\documentclass[a4paper,12pt]{article}

\usepackage{hijh}

\usepackage[normalem]{ulem}  
\usepackage{subcaption}
\usepackage{mathrsfs}

\newcommand{\De}{{\Delta}}
\newcommand{\Lm}{{\Lambda}}

\newcommand{\al}{{\alpha}}
\newcommand{\bt}{{\beta}}
\newcommand{\ga}{{\gamma}}
\newcommand{\de}{{\delta}}
\newcommand{\ep}{{\epsilon}}

\newcommand{\te}{{\theta}}
\newcommand{\ka}{{\kappa}}

\newcommand{\sig}{{\sigma}}
\newcommand{\vphi}{{\varphi}}

\newcommand{\bsC}{{\boldsymbol{C}}}

\newcommand{\bsE}{{\boldsymbol{E}}}
\newcommand{\bsG}{{\boldsymbol{G}}}
\newcommand{\bsR}{{\boldsymbol{R}}}

\newcommand{\bsh}{{\boldsymbol{h}}}

\newcommand{\bsW}{{\boldsymbol{W}}}
\newcommand{\bsPhi}{{\boldsymbol{\Phi}}}
\newcommand{\bsLm}{{\boldsymbol{\Lm}}}
\newcommand{\bsxi}{{\boldsymbol{\xi}}}
\newcommand{\bsV}{{\boldsymbol{V}}}
\newcommand{\bsT}{{\boldsymbol{T}}}

\newcommand{\cA}{{\mathcal{A}}}
\newcommand{\cB}{{\mathcal{B}}}

\newcommand{\cD}{{\mathcal{D}}}
\newcommand{\cE}{{\mathcal{E}}}
\newcommand{\cF}{{\mathcal{F}}}
\newcommand{\cK}{{\mathcal{K}}}
\newcommand{\cL}{{\mathcal{L}}}

\newcommand{\cV}{{\mathcal{V}}}
\newcommand{\cW}{{\mathcal{W}}}

\newcommand{\dal}{{\dot{\alpha}}}
\newcommand{\dbt}{{\dot{\beta}}}
\newcommand{\dga}{{\dot{\gamma}}}
\newcommand{\dmu}{{\dot{\mu}}}

\newcommand{\dg}{{\dagger}}

\newcommand{\lan}{{\langle}}
\newcommand{\nab}{{\nabla}}
\newcommand{\nn}{{\nonumber}}
\newcommand{\ol}{\overline}
\newcommand{\pd}{{\partial}}

\newcommand{\ran}{{\rangle}}

\def\Re{\mathop{\rm Re}\nolimits}
\def\Im{\mathop{\rm Im}\nolimits}

\def\bbra{{\langle\kern-2.5pt\langle}}
\def\kket{{\rangle\kern-2.5pt\rangle}}
\def\Bbra{{\Big\langle\kern-3.5pt\Big\langle}}
\def\Kket{{\Big\rangle\kern-3.5pt\Big\rangle}}

\begin{document}
\begin{center}
{\large A microscopic model for inflation from supersymmetry breaking}
\\
\medskip
\vspace{1cm}
\textbf{
I.~Antoniadis$^{\,a,b,}$\footnote{antoniadis@itp.unibe.ch}, 
A.~Chatrabhuti$^{c,}$\footnote{dma3ac2@gmail.com}, 
H.~Isono$^{c,}$\footnote{hiroshi.isono81@gmail.com}, 
R.~Knoops$^{c,}$\footnote{rob.k@chula.ac.th}
}
\bigskip

$^a$ {\small Laboratoire de Physique Th\'eorique et Hautes Energies - LPTHE,\\ 
Sorbonne Universit\'e, CNRS, 4 Place Jussieu, 75005 Paris, France }

$^b$ {\small Albert Einstein Center, Institute for Theoretical Physics,
University of Bern,\\ Sidlerstrasse 5, CH-3012 Bern, Switzerland }

$^c$ {\small Department of Physics, Faculty of Science, Chulalongkorn University,
\\Phayathai Road, Pathumwan, Bangkok 10330, Thailand }
\end{center}

\vspace{1cm}

\begin{abstract}

We have proposed recently a framework for inflation driven by supersymmetry breaking with the inflaton being a superpartner of the goldstino, that avoids the main problems of supergravity inflation, allowing for: naturally small slow-roll parameters, small field initial conditions, absence of a (pseudo)scalar companion of the inflaton, and a nearby minimum with tuneable cosmological constant. It contains a chiral multiplet charged under a gauged R-symmetry which is restored at the maximum of the scalar potential with a plateau where inflation takes place. The effective field theory relies on two phenomenological parameters corresponding to corrections to the K\"ahler potential up to second order around the origin. The first guarantees the maximum at the origin and the second allows the tuning of the vacuum energy between the F- and D-term contributions. Here, we provide a microscopic model leading to the required effective theory. It is a Fayet-Iliopoulos model with two charged chiral multiplets under a second ${\rm U}(1)$ R-symmetry coupled to supergravity. In the Brout-Englert-Higgs phase of this ${\rm U}(1)$, the gauge field becomes massive and can be integrated out in the limit of small supersymmetry breaking scale. In this work, we perform this integration and we show that there is a region of parameter space where the effective supergravity realises our proposal of small field inflation from supersymmetry breaking consistently with observations and with a minimum of tuneable energy that can describe the present phase of our Universe.

\end{abstract}

\newpage

\setcounter{tocdepth}{2}
\tableofcontents

\setcounter{footnote}{0}

\section{Introduction}
In a recent work~\cite{Antoniadis:2017gjr}, we have proposed a direct connection between inflation and supersymmetry breaking by identifying the 
inflaton with a superpartner of the Goldstone fermion of supersymmetry breaking (goldstino), charged under a gauged R-symmetry\footnote{See \cite{AlvarezGaume:2010rt,AlvarezGaume:2011xv,Ferrara:2016vzg} for earlier works on relating supersymmetry breaking with inflation and also \cite{Randall:1994fr,Riotto:1997iv,Izawa:1997jc,Buchmuller:2000zm,Schmitz:2016kyr,Domcke:2017xvu,Harigaya:2017jny,Domcke:2017rzu,Zheng:2016ftj,Antoniadis:2016aal,Antoniadis:2018} for related approaches along this direction.}. The superpotential is then linear in the inflaton superfield $X$ leading to a natural solution of the $\eta$-problem in supergravity\footnote{The $\eta$-problem is also evaded in hybrid inflation models by a somewhat similar way (see e.g.~\cite{Davali:1994}), but these models in general include several scalar fields besides the inflaton.}, due to an exact cancellation of the inflaton mass around the origin for canonically normalised kinetic terms, corresponding to a quadratic K\"ahler potential ${\cal K}=X{\bar X}+\dots$. A positive quartic correction to ${\cal K}$ is then needed to create a flat maximum at the origin providing naturally slow-roll small-field inflation in a model independent way. Indeed, the effective field theory has two parameters that can fit the amplitude and the spectral index of primordial density fluctuations, with a nice prediction for the number of e-folds and a rather small ratio of tensor-to-scalar perturbations.

The inflaton charge under the ${\rm U}(1)_{\rm R}$ should be small so that the D-term contribution to the scalar potential plays no role during inflation (thus driven by an F-term supersymmetry breaking) but could affect the minimum, allowing in particular for a tuning of the vacuum energy to an infinitesimal positive value. In order to study this question within the same effective field theory, an extra condition has to be imposed guaranteeing a `nearby' minimum that can be treated perturbatively around the maximum at the origin. It turns out that this is possible in the presence of a second order correction to the K\"ahler potential, cubic in $X{\bar X}$.

Obviously, an interesting question is whether the above desired corrections to the K\"ahler potential can arise from an underlying microscopic theory. In this work we provide an example of such a field theory model coupled to supergravity. It is the standard Fayet-Iliopoulos (FI) model of supersymmetric QED with a massive electron in the presence of a constant FI D-term \cite{Fayet:1974}. In global supersymmetry, there is a region of parameter space, when the FI parameter is large compared to the electron mass, where the ${\rm U}(1)$ is broken and supersymmetry breaking is dominated by an F-term but is still small compared to the ${\rm U}(1)$ mass. The spectrum is then approximately supersymmetric containing one massive vector multiplet and the light goldstino multiplet with a linear superpotential. The vector multiplet can be integrated out leading to an effective K\"ahler potential for the goldstino multiplet~\cite{Karim:2018}. The coefficient of the quartic term is, however, negative so that the origin is a minimum of the scalar potential upon coupling this model (naively) to supergravity.

In order to couple this model to supergravity, one has to promote the ${\rm U}(1)$ to a gauged R-symmetry. A mass term is therefore allowed only if the electron and positron have {\em not} opposite charges, since the superpotential has a net charge. Moreover, the FI parameter is fixed by the charge difference in terms of the Planck mass. In this work, we analyse this theory and show that there is a region in the parameter space where the ${\rm U}(1)$ is broken and the spectrum is approximately supersymmetric, so that the massive vector multiplet can be again integrated out leading to an effective K\"ahler potential for the goldstino multiplet. In this case, it turns out that the first order (quartic) correction can be positive so that the corresponding scalar potential has a maximum at the origin, providing a concrete example for the desired effective theory of the goldstino multiplet. Moreover, upon introducing a second ${\rm U}(1)$ (another gauged R-symmetry), we show that using the second order correction to the  K\"ahler potential one obtains a scalar potential describing a realistic inflation around the maximum with the  inflaton rolling down to a nearby minimum having a tuneable vacuum energy.

The outline of our paper is the following. In Section~2, we present for self-consistency a brief review of the proposed mechanism of inflation from supersymmetry breaking by identifying the inflaton with the goldstino superpartner. In Section~3, we consider the FI model based on an R-symmetry ${\rm U}(1)$ and  perform for illustration the integration out of the massive vector multiplet in global  supersymmetry ignoring the fact that this model is not consistent without supergravity. In Section~4, we perform the integration in supergravity using the superconformal formalism. In Section~5, we compute the effective field theory and we identify a region in the parameter space providing a realistic model for inflation with all desired properties. Section~6 contains our concluding remarks. For self-consistency and convenience for the reader, we have also an appendix with the  basic formalism of conformal supergravity that we use in Section~4.

\section{Inflation from Supersymetry breaking}
\label{sec:SUSY-inflation}

This section reviews a class of models studied recently by the present authors \cite{Antoniadis:2017gjr}, in which the inflaton is identified with the scalar superpartner of the goldstino in the presence of a gauged R-symmetry. The superpotential is then linear offering a natural solution to the $\eta$-problem. The K\"ahler potential is chosen such that inflation occurs in a plateau around the maximum of the scalar potential (hill-top), to avoid large field initial conditions, while the pseudoscalar partner of the inflaton is absorbed into the R-gauge field that becomes massive. Therefore, the inflaton is identified with the single scalar field that survives in the spectrum. Moreover, the model allows the presence of a realistic minimum with an infinitesimal positive vacuum energy. This is realised due to a cancellation between the F- and D-term contributions to the scalar potential, without affecting the properties of the inflationary plateau.  

In general, such models can be classified into two classes depending on whether the maximum corresponds to a point of unbroken (case 1) or broken (case 2) R-symmetry. In the following, we will summarise the main features of models of case 1 that we consider in this work, where inflation occurs near the maximum of the scalar potential where R-symmetry is restored.

Let us consider supergravity theories containing a single chiral multiplet transforming under a gauged R-symmetry with a corresponding abelian vector multiplet.  We assume that the chiral multiplet $X$ transforms as:
\begin{align}
X \rightarrow e^{-i q \Lambda} X,
\end{align}
where $q$ is the charge of $X$, and $\Lambda$ is the gauge parameter. The K\"ahler potential is therefore a function of $X\bar X$ while the superpotential $\mathcal{W}$ is linear in $X$ 
\begin{align}
\mathcal{K} = \mathcal{K}(X \bar X), \quad \mathcal{W} = \kappa^{-3} f X,
\end{align}
where $f$ is a constant.  Note that $X$ is dimensionless and $\kappa^{-1} = 2.4\times 10^{18}$ GeV is the reduced Planck mass.  The gauge kinetic function is taken to be 1.  Note that the superpotential is not gauge invariant under the ${\rm U}(1)$ gauge symmetry, but it transforms as $\mathcal{W} \rightarrow \mathcal{W}e^{-iq \Lambda}$. Therefore, the ${\rm U}(1)$ is a gauged R-symmetry which in Section \ref{sec:eff-inflation} we will  denote by ${\rm U}(1)^\prime$.

The scalar potential is given by
\begin{align}
\mathcal{V} &= \mathcal{V}_F + \mathcal{V}_D,\\
\mathcal{V}_F &= e^{\kappa^2\mathcal{K}}\left( -3\kappa^2 \mathcal{W} \bar{\mathcal{W}} + \nabla_X \mathcal{W} g^{X\bar X}\bar{\nabla}_{\bar X}\bar{\mathcal{W}}\right),\\
\mathcal{V}_D &= \frac{1}{2}\mathcal{P}^2, \label{Dpot}
\end{align}
where the K\"ahler covariant derivative is acting on $\mathcal{W}$ as
\begin{align}
\nabla_X \mathcal{W} = \partial_X \mathcal{W}(z) + \kappa^2 (\partial_X\mathcal{K})\mathcal{W}.
\end{align}
The moment map $\mathcal{P}$ is given by
\begin{align}
\mathcal{P} = i(k^X\partial_X\mathcal{K} - r).
\end{align}
where $k^X$ is the Killing vector for $X$ under the ${\rm U}(1)$ R-symmetry, and $r$ is defined by $r=-\ka^{-2}k^X\cW_X/\cW$; in the present setup, they become $k^X = -iqX, ~ r=i\ka^{-2}q$. As usual, subscripts stand for partial derivatives: $\cW_X:=\partial_X\cW$.

We are interested in the case where inflation starts near a local maximum of the potential at $X = 0$, where R-symmetry is preserved. Let us expand the K\"ahler potential in $X\bar X$ up to quadratic order:
\begin{align}
\kappa^2\mathcal{K} =  X\bar X + A (X \bar X)^2 + ... \, .
\end{align} 
With this, the F-term potential becomes
\begin{align}
\kappa^{4}\mathcal{V}_F =  f^2 e^{X \bar X \left( 1 + A X \bar X \right) }
 \left[-3 X \bar X + \frac{\left( 1 + X \bar X  + 2 A (X \bar X)^2) \right)^2}{1 + 4 A X \bar X} \right], 
\end{align}
and the D-term potential is
\begin{align}
  \kappa^{4}\mathcal V_D =  \frac{q^2}{2} \left[ 1 + X \bar X  + 2 A (X \bar X)^2 \right]^2 .
\end{align}
Under a change of field variables 
\begin{align}
X = \rho e^{i \theta}, \quad \bar X = \rho e^{-i \theta}, \quad (\rho\ge 0),
\end{align}
the scalar potential reads
\begin{align}
 \kappa^4 \mathcal V = f^2  e^{\rho^2 + A \rho^4}  \Big[ - 3 \rho^2 + \frac{\left( 1 + \rho^2 + 2 A \rho^4\right)^2}{1 + 4 A \rho^2} \Big] + \frac{q^2}{2} \left( 1 + \rho^2 + 2 A \rho^4 \right)^2 . \label{scalarpot}  
\end{align}
Note that the scalar potential is only a function of the modulus $\rho$ and the phase $\theta$ will be ``eaten" by the ${\rm U}(1)_{\rm R}$ gauge field in a similar manner to the standard Brout-Englert-Higgs mechanism.

We now interpret the field $\rho$ as the inflaton. In order to calculate the slow-roll parameters, one needs to work with the canonically normalised field $\chi$ satisfying
\begin{align}
 \frac{d\chi}{d\rho} = \sqrt{2 {\cal K}_{X \bar X} }. \label{case1_norm}
\end{align}
The slow-roll parameters are given in terms of the canonical field $\chi$ by
\begin{align}
\epsilon = \frac{1}{2\kappa^2} \left( \frac{d\mathcal V/d\chi}{\mathcal V}\right)^2, \quad
 \eta = \frac{1}{\kappa^2} \frac{d^2\mathcal V/d\chi^2}{\mathcal V}. 
 \label{slowroll_pars2}
\end{align}
Since we assume inflation to start near $\rho=0$, we expand  
\begin{align}
 \epsilon &=  4 \left( \frac{-4 A  + y^2}{2 + y^2} \right)^2 \rho^2
 + \mathcal O(\rho^4),  \notag \\
 \eta &= 2 \left( \frac{-4 A  +  y^2}{2  + y^2} \right) + \mathcal O(\rho^2) \label{case1_slowroll} ,
\end{align}
where we defined $y = q/f$.  
The above equation implies $\epsilon\simeq\eta^2\rho^2 \ll \eta$.   For simplicity, we consider the case where the F-term potential is dominant by setting $y$ to be very small so that $y$ can be neglected. Taking this into account, let us find some constraints on the coefficient $A$ of the quadratic term of the K\"ahler potential. The condition that the scalar potential has a local maximum at $\rho=0$ requires $A>0$. Furthermore, the slow-roll condition $|\eta| \ll 1$ gives an upper bound $A \ll 0.25$. Therefore, the constraint on $A$ is
\begin{align}
0<A \ll 0.25.
\label{const_A}
\end{align}       
In order to satisfy CMB observational data with $\eta \sim -0.02$, we choose $A \sim 0.005$.  In the following sections we explore a microscopic model that can generate the coefficient $A$ satisfying the requirement (\ref{const_A}).

\section{Fayet-Iliopoulos model in global supersymmetry}
\label{sec:FI-global}
In this section, we introduce a ``generalisation" of the Fayet-Iliopoulos model as an example of the microscopic origin for the effective field theory of the inflation model described in the previous section. We consider the regime where both gauge symmetry and supersymmetry are spontaneously broken, leaving (in the decoupling limit) the goldstino as the only light mode in this sector.  
\subsection{Setup}
We consider a globally supersymmetric theory specified by the following K\"ahler potential, superpotential and gauge kinetic function
\begin{align}
\mathcal{K} &= \bar\bsPhi_+ e^{q_+\bsV} \bsPhi_+ + \bar\bsPhi_- e^{-q_-\bsV} \bsPhi_-, \\
\mathcal{W} &= m\bar\bsPhi_+\bsPhi_-, \\
\mathcal{F} &= 1+b\ln\frac{\bsPhi_-}{M},
\end{align}
where $\bsV$ is the vector superfield associated with the gauged ${\rm U}(1)$ transformation.   $M$ is a mass scale parameter which will be fixed later. In this globally supersymmetric model we let the fields and parameters be dimensionful. The two chiral multiplets $\bsPhi_\pm$ and the vector superfield transform under the gauge transformation as
\begin{align}
\bsPhi_\pm \mapsto e^{\mp iq_{\pm}\bsLm}\bsPhi_\pm, \qquad
\bsV \mapsto \bsV+i(\bsLm-\bsLm^\dg),
\label{gauge_tranf}
\end{align}
where $\bsLm$ is a gauge parameter chiral superfield.
The logarithmic term in the gauge kinetic function is needed to cancel a chiral anomaly in the case $q_+ \neq q_-$ with an appropriate coefficient $b$~\cite{Antoniadis:2014iea}.  Note that the case of $q_+ = q_-$ is studied in \cite{Karim:2018}. In our case, $\mathcal{W}$ is not invariant under \eqref{gauge_tranf} and ${\rm U}(1)$ is an R-symmetry. The action we consider is given by
\begin{align} 
S &= \frac{1}{4}\int \! d^4x ~ [\mathcal{F}(\bsPhi_-)\bsW\bsW]_{\te\te} + {\rm h.c.}\nn\\
&\quad+ \int \! d^4x ~ [m\bsPhi_+\bsPhi_-]_{\te\te} + {\rm h.c.}\nn\\
&\quad+ \int \! d^4x \, [\bar\bsPhi_+ e^{q_+\bsV} \bsPhi_+ + \bar\bsPhi_- e^{-q_-\bsV} \bsPhi_- + \xi q_-\bsV]_{\te\te\bar\te\bar\te},
\end{align}
where we introduced the FI parameter $\xi$ of mass dimension 2. 

Note that gauging the R-symmetry is not consistent in global supersymmetry. However, as we mentioned in the introduction, we ignore this problem and consider the above model for illustration of the integration out procedure, as a warming up exercise, before going to supergravity.

\subsection{Mass spectrum}
We first investigate the mass spectrum of the theory. For this we adopt the Wess-Zumino gauge. Note that the auxiliary fields enter the superfields as
\begin{align}
\bsPhi_\pm \ni \te\te F_\pm, \qquad
\bsV \ni \frac{1}{2}\te\te\bar\te\bar\te D.
\end{align}
The part of the action with auxiliary fields reads
\begin{align}
S ~ \ni & ~
\int \! d^4x ~ \frac{1}{4}(2+b\ln|\vphi_-/M|^2)D^2
\nn\\
&\quad
+ \int \! d^4x \, 
\left(
\frac{1}{2}q_+D|\vphi_+|^2 - \frac{1}{2}q_-D|\vphi_-|^2
+ \bar F_+F_+ + \bar F_-F_- + \frac{1}{2}\xi q_-D
\right) 
\nn\\
&\quad
+ \int \! d^4x ~ m(F_+\vphi_- + F_-\vphi_+ + \bar F_+\bar\vphi_- + \bar F_-\bar\vphi_+).
\end{align}
After integrating out $F_\pm$ and $D$, this becomes
\begin{align}
S ~ \ni & ~
- \int \! d^4x \, \frac{1}{4} 
\frac{\left( q_+|\vphi_+|^2 - q_-|\vphi_-|^2 + \xi q_- \right)^2}{2+b\ln|\vphi_-/M|^2}
- \int \! d^4x ~ m^2(|\vphi_-|^2 + |\vphi_+|^2)\, ,
\end{align}
leading to the scalar potential
\begin{align}
\mathcal{V}^{\rm {UV}} = 
\frac{1}{4} 
\frac{\left( q_+|\vphi_+|^2 - q_-|\vphi_-|^2 + \xi q_- \right)^2}{2+b\ln|\vphi_-/M|^2}
+ m^2(|\vphi_-|^2 + |\vphi_+|^2).
\end{align}
We are interested in the spectrum around the vacuum (for $\xi>0$)
\begin{align}
\lan \vphi_+ \ran = 0, \qquad \lan \vphi_- \ran = v.
\end{align}
To simplify the expressions, we may take the scale parameter $M = v$, getting rid of the factor $\ln v$. The vacuum expectation values of the auxiliary fields are
\begin{align}
\lan F_+ \ran = 0, \qquad \lan F_- \ran = -mv, \qquad
\lan D \ran = -\frac{q_-}{2}(\xi - v^2).
\end{align}
For our convenience, let us introduce new parameters 
\begin{align}\label{deltadef}
\Delta := \xi - v^2 ~\text{ and }~x := \frac{q_+}{q_-}.
\end{align}
Since the first derivative of the potential $\mathcal{V}^{\rm {UV}}_{\vphi_-^*}$ must vanish at $\vphi_-=v$, this gives us the constraint equation
\begin{align}
-\frac{1}{4}q_-^2 v^2\Delta
-\frac{1}{16}bq_-^2 \Delta^2
+m^2v^2 = 0.
\label{global_const_eq}
\end{align}

It would be clearer for the reader to start our discussion with the approximation of $b=0$ where (\ref{global_const_eq}) has a unique solution,
\begin{align}
\Delta = \frac{4m^2}{q_-^2}.
\label{Delta_global}
\end{align}
One can easily see that the imaginary part $\Im \vphi_-$ plays the role of R-Goldstone boson and is eaten by the ${\rm U}(1)$ gauge field.  The real part $\Re\vphi_-$  has mass $|q_-|v$ (the same as the ${\rm U}(1)$ gauge field) while $\vphi_+$ has mass square $m^2(1+x)$.  In the next subsection we will integrate out the massive vector multiplet and leave only $\bsPhi_+$ in the low-energy effective theory.  This can be done consistently if the parameter $m$, $v$ and $q_-$ satisfy the following integrating out condition,
\begin{align}
m^2 \ll q_-^2v^2  ~~\text{ or } ~~ \Delta \ll v^2. 
\label{global_const}
\end{align}

For the more general case where $b \neq 0$, equation (\ref{global_const_eq}) becomes quadratic and has two solutions $\De=\De_\pm$ where 
\begin{align}
\Delta_\pm = -\frac{2v^2}{b} \pm \frac{2v^2}{b}\sqrt{1+\frac{4bm^2}{q_-^2v^2}}.
\label{Del_sol_global}
\end{align}
For small $m$, they become
\begin{align}
\Delta_\pm = \left\{\begin{array}{c} 4m^2/q_-^2 \\ -4m^2/q_-^2-4v^2/b \end{array}\right. .
\end{align}
The mass of $\vphi_+$ is determined by the second derivative of the potential with respect to $\vphi_+$ and $\vphi_+^*$,
\begin{align}
m^2_{\varphi_+}=\mathcal{V}^{\rm {UV}}_{\vphi_+^*\vphi_+}|_{\rm vac} &= 
m^2+\frac{1}{4}q_+q_- \Delta.
\end{align}
The second derivatives of the potential with respect to $\vphi_-$ are:
\begin{align}
\cV^{\rm {UV}}_{\vphi_-^*\vphi_-^*}|_{\rm vac} &= 
\frac{1}{4}q_-^2 v^2
+ \frac{1}{4}bq_-^2 \Delta
+ \frac{1}{16}b(b+1)q_-^2 \frac{\Delta^2}{v^2}, \\
\cV^{\rm {UV}}_{\vphi_-^*\vphi_-}|_{\rm vac} &= 
m^2+\frac{1}{4}q_-^2 v^2
+ \frac{1}{4}(b-1)q_-^2 \Delta
+ \frac{1}{16}b^2q_-^2 \frac{\Delta^2}{v^2}.
\end{align}
One can easily see that in the region of parameter space where the integrating out constraint (\ref{global_const}) is satisfied, $\vphi_+$ is the lightest field.

\subsection{Integrating out heavy fields in superspace}
We adopt the unitary gauge $\Phi_-=v$, for which the gauge parameter is
\begin{align}
\bsLm = -\frac{i}{q_-}\ln\frac{v}{\bsPhi_-}.
\label{unitary_gauge}
\end{align}
The action in this gauge reads
\begin{align}
S &= 
\frac{1}{4}\int \! d^4x ~ [\bsW\bsW]_{\te\te} + {\rm h.c.}
\nn\\
&\quad
+ \int \! d^4x ~ mv[\bsPhi_+]_{\te\te} + {\rm h.c.}
\nn\\
&\quad
+ \int \! d^4x \, 
[\bar\bsPhi_+ e^{xq_-\bsV} \bsPhi_+ + v^2 e^{-q_-\bsV} + \xi q_-\bsV]_{\te\te\bar\te\bar\te}.
\end{align}

We now integrate out $\bsV$ around its vacuum. 
Its equation of motion is
\begin{align}
\label{globaleomV}
\frac{1}{4}\cD\bar\cD^2\cD\bsV + xq_-\bar\bsPhi_+ e^{xq_-\bsV} \bsPhi_+ - q_-v^2e^{-q_-\bsV} + \xi q_- = 0.
\end{align}
Due to the FI term $q_-\xi$, the vacuum solution cannot be $\bsV=0$, but its highest component $D$ acquires a non-zero vacuum expectation value.
To remove the tadpole, we make a shift and introduce the (fluctuating superfield) variable $\hat\bsV$ around the vacuum~\cite{Karim:2018},
\begin{align}
\bsV=\hat\bsV+\frac{1}{2}\te\te\bar\te\bar\te\lan D \ran, \qquad \lan D \ran = -\frac{q_-}{2}\Delta,
\label{def_Vhat}
\end{align}
and the equation of motion becomes
\begin{align}
\label{globaleomhatV}
\frac{1}{4}\cD\bar\cD^2\cD\hat\bsV + xq_-\bar\bsPhi_+ e^{xq_-\bsV} \bsPhi_+ + q_-v^2(1-e^{-q_-\bsV}) = 0.
\end{align}
To integrate out heavy degrees of freedom at tree level with supersymmetry kept, we neglect the derivative term $\cD\bar\cD^2\cD\hat\bsV$, to find the low energy effective equation of motion
\begin{align}
\label{globaleomnohatV}
xq_-\bar\bsPhi_+ e^{xq_-\bsV} \bsPhi_+ + q_-v^2(1-e^{-q_-\bsV}) \simeq 0.
\end{align}
This gives us the following relation 
\begin{align}
\bar\bsPhi_+\bsPhi_+ =x^{-1}{v ^2 e^{- q_- \bsV (x+1)} (1-e^{q_- \bsV} )}.
\label{globalSol_Phi}
\end{align}

Let us now integrate $\bsV$ in the action. For this, we first rewrite the $\bsV$-dependent part in the action as
\begin{align}
\label{globalactionnoDDDD}
&\quad
\int d^4xd^4\te  \left[
\frac{1}{8}\bsV\cD\bar\cD^2\cD \bsV + \bar\bsPhi_+ e^{xq_-\bsV} \bsPhi_+ + v^2 e^{-q_-\bsV} + \xi q_-\bsV
\right]
\nn\\
&=
\int d^4xd^4\te \Big[
-\frac{1}{2}\bsV\left(xq_-\bar\bsPhi_+ e^{xq_-\bsV} \bsPhi_+ - q_-v^2e^{-q_-\bsV} + \xi q_-\right) 
\nn\\
&\qquad\qquad\qquad
+ \bar\bsPhi_+ e^{xq_-\bsV} \bsPhi_+ + v^2 e^{-q_-\bsV} + \xi q_-\bsV
\Big],
\end{align}
where in the second line we applied the equation of motion \eqref{globaleomV}.
Using the relation \eqref{globalSol_Phi} in the action \eqref{globalactionnoDDDD}, we can derive the effective K\"ahler potential for the light goldstino superfield $\bsPhi_+$ in the global supersymmetry (SUSY) case as
\begin{equation}
\mathcal{K}_{\rm eff} = -\frac{v^2}{x} + \frac{q_-  \left(v^2+\xi \right)\bsV}{2}+\frac{v^2 (x+1)e^{-q_- \bsV}}{x},
\label{Keff_global}
\end{equation}
where $\bsV$ must be understood as a function of $\bar\bsPhi_+\bsPhi_+$ by inverting equation~\eqref{globalSol_Phi}.

\subsection{The effective K\"ahler potential near the maximum of the scalar potential}
\label{subsec:inflation-global}

In this subsection we explore the behaviour of the effective K\"ahler potential (\ref{Keff_global}) near the maximum of the scalar potential at $\bsPhi_+ = 0$.  For simplicity, let us absorb $q_-$ into the vector multiplet by rescaling $q_-\bsV \rightarrow \bsV$.  Since $\bsV$ can not be expressed explicitly in terms of $\bsPhi_+$,  we consider its expansion 
\begin{equation}
\label{pert_sol_global}
\bsV = V_0 + V_1\bar\bsPhi_+\bsPhi_+ + V_2 (\bar\bsPhi_+\bsPhi_+)^2 + V_3 (\bar\bsPhi_+\bsPhi_+)^3  +  ... ~.
\end{equation}
Using equation \eqref{globaleomnohatV} we obtain the solution
\begin{eqnarray}
V_0 &=& 0, ~ V_1 = -\frac{x}{v^2} ,~ V_2 = \frac{x^2 (2 x+1)}{2 v^4},\nonumber\\
V_3 &=& -\frac{x^3 (9 x (x+1)+2)}{6 v^6}.
\end{eqnarray}
Substituting this back into the effective K\"ahler potential \eqref{Keff_global}, we obtain
\begin{eqnarray}
\kappa^2 \mathcal{K}_{\rm eff} &=&v^2 + \left( 1-\frac{x}{2v^2}\Delta\right)\bar\bsPhi_+\bsPhi_+ -\frac{x^2 \left(2v^2- \Delta(2x+1) \right)}{4 v^4}|\bar\bsPhi_+\bsPhi_+|^2  \nonumber\\
 && + \frac{x^3 (3 x+1) \left(2v^2 - \Delta(3x+2) \right)}{12 v^6}|\bar\bsPhi_+\bsPhi_+|^3 + ... \, .
\label{Keff_expandglobal}
\end{eqnarray}

In order to make a comparison with the previous section, we  define the canonically normalised chiral superfield $\bsPhi$ as
\begin{equation}
\label{chiral}
\bsPhi :=  \sqrt{1-\frac{x}{2v^2}\Delta}\;\bsPhi_+.
\end{equation} 
The constant term in \eqref{Keff_expandglobal} can be absorbed by a K\"ahler transformation. Then, the effective K\"ahler potential can be written as
\begin{equation}
 \mathcal{K}_{\rm eff} =  |\bar{\bsPhi}\bsPhi| +A_2 |\bar{\bsPhi}\bsPhi|^2 + A_3 |\bar{\bsPhi}\bsPhi|^3  + ... \, ,
 \label{Keff}
\end{equation}
where 
\begin{eqnarray}
A_2 & = & -\frac{x^2 \left(2v^2-\Delta(2x+1)\right)}{\left(2v^2 -  x\Delta\right)^2},\label{A2_global} \label{A2_global}\\
A_3 & = & \frac{2 x^3 (3 x+1) \left(2v^2-\Delta  (3 x+2)\right)}{3 \left(2v^2 -  x\Delta \right)^3}.
\label{A3_global}
\end{eqnarray}
The condition that the scalar potential has a local maximum at the origin requires that $A_2 > 0$.  From \eqref{A2_global}, this requirement implies that $\Delta > v^2$, which violates the integrating out condition (\ref{global_const}).  In the following sections, we will show that this problem can be avoided by taking the supergravity effect into account.

\section{Fayet-Iliopoulos model in supergravity}
In the UV model of the last section, the gauged ${\rm U}(1)$ transformation changes the superpotential, being an R-transformation. Since it is gauged, it  involves a local phase rotation of the fermionic coordinates of superspace. This forces us to resort to supergravity.

In this section, we first present a supergravity extension of the generalised FI model with two chiral multiplets $\bsPhi_\pm$ and one vector multiplet.
This theory also has a vacuum in which only $\bsPhi_+$ is lighter than the other degrees of freedom. We then integrate out the heavy degrees of freedom to find an effective supergravity action in $\bsPhi_+$.  In the next section, we will consider its applications, showing that this model avoids the problem mentioned at the end of the last section.

\subsection{UV action}
The UV supergravity action we consider is
\begin{align}
S &= 
\frac{1}{4}\int \! d^4xd^2\te ~ \boldsymbol{\cE} \cF(\bsPhi_-)\bsW^\al\bsW_\al + {\rm h.c.}
\nn\\
&\quad
+ \kappa^{-3} m \int \! d^4xd^2\te ~ \boldsymbol{\cE}\bsPhi_+\bsPhi_- + {\rm h.c.}
\nn\\
&\quad
- 3\kappa^{-2} \int \! d^4xd^4\te \, 
\bsE e^{-\kappa^2\cK_0/3 - (q_+-q_-)\bsV/3},
\label{matterYM-sugra}
\end{align}
which is formulated in Poincar\'e superspace as in \cite{Wess:1992cp}. This theory is invariant under a gauged ${\rm U}(1)$ transformation which acts only on matter superfields,
which we call ${\rm U}(1)_{\rm m}$ transformation. 
In the following we will make all fields and parameters dimensionless in the unit of the reduced Planck mass $\ka^{-1}$, in contrast to the last section.

The notation is as follows: 
The chiral superfields $\bsPhi_\pm$ transform under ${\rm U}(1)_{\rm m}$ as,\footnote{Strictly speaking, it involves a local rotation of the fermionic coordinates in addition to the overall phase rotation, due to the non-invariance of the superpotential under ${\rm U}(1)_{\rm m}$. In this sense, ${\rm U}(1)_{\rm m}$ is a gauged R-transformation, which is allowed only in supergravity.}
\begin{align}
\bsPhi_\pm \mapsto e^{\mp iq_\pm\bsLm}\bsPhi_\pm, \qquad
\end{align}
where $\bsLm$ is chiral.
The vector superfield $\bsV$ transforms under ${\rm U}(1)_{\rm m}$ as
\begin{align}
\bsV \mapsto \bsV+i(\bsLm-\bar\bsLm).
\end{align}
The function $\cK_0$ is the ${\rm U}(1)_{\rm m}$-invariant K\"ahler potential,
\begin{align}
\kappa^2\cK_0 = \bar\bsPhi_+ e^{q_+\bsV} \bsPhi_+ + \bar\bsPhi_- e^{-q_-\bsV} \bsPhi_-,
\end{align}
and $\bsW_\al$ is the gaugino superfield, defined with the super-Poincar\'e covariant derivatives $\cD_\al,\bar\cD_\dal$ as
\begin{align}
\bsW_\al = -\frac{1}{4}\bar\cD^2\cD_\al \bsV.
\end{align}
The function $\cF(\bsPhi_-)$ is the gauge kinetic function, given by 
\begin{align}
\cF(\bsPhi_-) = 1+b\ln\bsPhi_-, \qquad b=\frac{(x-1)^3q_-^2}{24\pi^2},
\end{align}
in which the second term produces a Green-Schwarz action that cancels the chiral anomaly of ${\rm U}(1)_{\rm m}$. For more details see \cite{Antoniadis:2017gjr,Antoniadis:2014iea}.

The scalar potential of the UV theory \eqref{matterYM-sugra} is given by
\begin{align}
\kappa^4 \cV^{\rm {UV}} &= \frac{1}{4}q_-^2\frac{\big(x|\varphi_+|^2-|\varphi_-|^2 + x - 1 \big)^2}{2(1+b\ln v)} \nn\\
&\qquad + m^2 e^{|\varphi_+|^2 + |\varphi_-|^2}\big(|\varphi_+|^2 + |\varphi_-|^2 -|\varphi_+|^2|\varphi_-|^2 \big),
\label{scpot-U(1)}
\end{align}
where $\vphi_\pm=\bsPhi_\pm|$ is the lowest component of superfields $\bsPhi_\pm$. The first line is the D-term contribution. Note that it contains the Fayet-Iliopoulos type contribution with FI parameter $x-1$.  As a result, in the supergravity case, it is natural to introduce the parameter $\De$ as
\begin{align}
\De:=x-1-v^2.
\label{Delta_sugra}
\end{align} 
As in the last section, we are interested in a vacuum of the form
\begin{align}
\langle\vphi_+\rangle=0, \qquad \langle\vphi_-\rangle=v,
\end{align}
which spontaneously breaks ${\rm U}(1)_{\rm m}$ and supersymmetry and around which the fields of $\bsV,\bsPhi_-$ are heavier than those of $\bsPhi_+$, in the limit of small SUSY breaking scale. 
The extremisation condition with respect to $\vphi_-$ reads
\begin{align}
-\frac{1}{4}q_-^2 \frac{\De}{1+b\ln v} v^2
-\frac{1}{16}bq_-^2 \left(\frac{\De}{1+b\ln v}\right)^2
+m^2v^2(1+v^2)e^{v^2} = 0.
\label{determine_q}
\end{align}
This gives us a constraint among the parameters $\Delta$, $v$, $x$ and $q_-$  which will be used in Section \ref{sec:eff-inflation}.

We can consider first the approximation $b=0$.  In this case equation (\ref{determine_q}) has a unique solution
\begin{align}
\De = \frac{4m^2}{q_-^2}(1+v^2)e^{v^2}.
\label{Delta_local}
\end{align}
It is also easy to see that $\Im\vphi_-$ is still the massless R-Goldstone boson while $\Re\vphi_-$ gets a correction to its mass-squared compared to the global SUSY case $q_-^2v^2$ by $4m^2v^2(2+v^2)e^{v^2}$.  The mass-squared of $\vphi_+$ also changes to $m^2(1+x+xv^2)e^{v^2}$ and the integrating out condition is satisfied if this mass is much smaller than the other masses.  

For $b\neq 0$ eq.~(\ref{determine_q}) gives two solutions $\De=\De_\pm$, where $\De_\pm$ are given by
\begin{align}
\Delta_\pm := \frac{2v^2(1+b\ln v)}{b} \bigg( \! -1 \pm \sqrt{1+\frac{4bm^2(1+v^2)e^{v^2}}{q_-^2v^2}} ~ \bigg).
\end{align}
Notice that the existence of the two solutions originates from the anomaly coefficient $b$. 
The mass$^2$ of the vector field $A_\mu$ is $q_-^2v^2$.
The mass matrices of $\vphi_\pm$ are given by
\begin{align}
\cV^{\rm {UV}}_{\vphi_+^*\vphi_+^*}|_{\rm vac} &= 0, \\
\cV^{\rm {UV}}_{\vphi_+^*\vphi_+}|_{\rm vac} &= 
m^2e^{v^2}+\frac{1}{4}xq_-^2 \frac{\Delta}{1+b\ln v}, \label{mass_phi_p}\\
\cV^{\rm {UV}}_{\vphi_-^*\vphi_-^*}|_{\rm vac} &= 
m^2e^{v^2}v^2(2+v^2)
+ \frac{1}{4}q_-^2 \frac{v^2}{1+b\ln v}
+ \frac{1}{16}bq_-^2 \frac{\Delta^2}{v^2(1+b\ln v)^2} \nn\\
&\quad
+ \frac{1}{4}bq_-^2 \frac{\Delta}{(1+b\ln v)^2}
+ \frac{1}{16}b^2q_-^2 \frac{\Delta^2}{v^2(1+b\ln v)^3}, \label{mass_phi_m2}\\
\cV^{\rm {UV}}_{\vphi_-^*\vphi_-}|_{\rm vac} &= 
m^2e^{v^2}(1+3v^2+v^4)
+ \frac{1}{4}q_-^2 \frac{v^2}{1+b\ln v}
- \frac{1}{4}q_-^2 \frac{\Delta}{1+b\ln v}
 \nn\\
&\quad
+ \frac{1}{4}bq_-^2 \frac{\Delta}{(1+b\ln v)^2}
+ \frac{1}{16}b^2q_-^2 \frac{\Delta^2}{v^2(1+b\ln v)^3}. \label{mass_phi_m1}
\end{align}
In this section, we assume that the integrating out procedure is justified, which we will show explicitly in Section~\ref{sec:eff-inflation} with the analysis of the parameter space leading to models of realistic inflation.

\subsection{Normalisation, compensators, and conformal supergravity}
\label{subsec:confsugra}
The next task is to integrate out the heavy degrees of freedom and to identify the resulting effective K\"ahler potential and superpotential.
In general, the form of the effective action highly depends on the normalisation of the kinetic terms of the gravity multiplet in the UV theory. In this subsection, we discuss how to control the normalisation dependence in the effective theory and propose a method of choosing the normalisation which facilitates the identification of the effective K\"ahler potential and superpotential and the computation of the scalar potential.

\subsubsection*{Normalisation and compensator}
The supergravity action coupled to matter is specified by a K\"ahler potential $\cK$ and a superpotential $\cW$
\cite{Wess:1992cp}:
\begin{align}
-3\ka^{-2} \int \! d^4xd^4\te \, \bsE e^{-\ka^2 \cK/3}.
\end{align}
In components, the kinetic terms of the gravity multiplet take the following form:
\begin{align}
e^{-\ka^2 \cK/3}| \times({\rm canonical~one}),
\end{align} 
where the symbol $|$ picks up the lowest component.
We can control the normalisation by rescaling the gravity multiplet.
This may be performed in components \cite{Wess:1992cp}, but in this article we will do it in superspace to keep supersymmetry manifest. 
We recall that 
\begin{align}
-3\ka^{-2} \int \! d^4xd^4\te \, \bsE
\label{can-dtype}
\end{align}
gives the canonically normalised kinetic terms in  the gravity multiplet.

A way to make manifest the Weyl rescaling of the metric in superspace is to introduce ``compensator'' superfields along with additional local transformations $G_c$. The new action with the compensators is defined to be invariant under $G_c$ in addition to the super-diffeomorphism/local super-Poincar\'e invariance. We illustrate this below.

Given a supergravity action $S_0[\{\bsPhi\}]$ in Poincar\'e superspace, we define a new action $S[\{\bsC\};\{\bsPhi\}]$ with compensators $\{\bsC\}$.
The $G_c$ invariance is then dictated by\footnote{Normally, matter chiral superfields $\bsPhi$ and the vector superfield $\bsV$ are taken to be invariant. On the other hand, the vierbein $\bsE$ and the gaugino superfield $\bsW_\al$ transform under $G_c$.}
\begin{align}
S[\{\bsC\};\{\bsPhi\}] = S[\{\bsC'\};\{\bsPhi'\}],
\end{align}
where $G_c$  induces transformations $\bsPhi \mapsto \bsPhi'$, $\bsC \mapsto \bsC'$.
We can recover the original action by gauge fixing $\{\bsC\}$ to 1, exhausting $G_c$ degrees of freedom,
\begin{align}
S_0[\{\bsPhi_0\}]=S[\{1\};\{\bsPhi_0\}].
\end{align}
On the other hand, an action $S_{\rm can}[\{\bsPhi\}]$ with canonically normalised kinetic terms can also be realised by another gauge fixing $\bsC=\bsC_{\rm can}$ that exhausts $G_c$,
\begin{align}
S_{\rm can}[\{\bsPhi_{\rm can}\}]=S[\{\bsC_{\rm can}\};\{\bsPhi_{\rm can}\}].
\end{align}
These actions are physically equivalent since $G_c$ is gauged.
Note that depending on $G_c$, we need to enlarge the geometry (namely, modify the covariant derivatives) of the superspace over which $S[\{\bsC\};\{\bsPhi\}]$ is defined, as we will see shortly.

\medskip

In this article we will consider the case where the compensators are $\bsC,\ol\bsC$ which enter the D-term action as
\begin{align}
-3\ka^{-2} \int \! d^4xd^4\te \, \bsE\bsC\ol\bsC e^{-\ka^2 \cK/3}.
\label{D-type}
\end{align}

\subsubsection*{Superconformal transformations as $\boldsymbol{G_c}$ and conformal supergravity}
A simple choice of $G_c$ is the super-Weyl transformation \cite{Kaplunovsky:1994fg}, which changes the compensators as
\begin{align}
\bsC \mapsto e^{-2\bsLm}\bsC,
\end{align}
where the underlying superspace is the Poincar\'e superspace \cite{Wess:1992cp} and $\bsC,\bsLm$ are chiral. The canonical normalisation of the gravity multiplet may be realised by the choice $\bsC|=e^{\ka^2\cK/6}|$. The choice $\bsC=e^{\ka^2\cK/6}$ however, which would realise \eqref{can-dtype}, is not allowed since it is not chiral and thus breaks supersymmetry.

In this article, we take another option, in which $G_c$ is large enough that it allows the gauge fixing
\begin{align}
& \bsC=\ol\bsC=e^{\ka^2\cK/6}, \label{canonical-fixing1}
\end{align} 
leading, as we will see later, to a simple gauge-fixed action that facilitates the identification of the effective K\"ahler potential and superpotential.
Among several choices of $G_c$ proposed along this line, we adopt the one used by Butter \cite{Butter:2009cp}, which is generated by the dilatation, chiral ${\rm U}(1)$ rotation, and special conformal transformations;\footnote{Note that these transformations are internal.}
namely, the super-Poincar\'e transformations plus $G_c$ form the superconformal ones.
Since $G_c$ is gauged, we introduce the (superconformal) covariant derivatives with the gauge fields for $G_c$.
A superspace with these covariant derivatives is called the conformal superspace.

\subsubsection*{Gauge fixing of compensators}
Butter \cite{Butter:2009cp} presented a formalism of the conformal superspace and supergravity actions over it  (conformal supergravity) with compensators, and exemplified their relations to other formulations of supergravity. In particular, he proved that for a given function $\cK$, the gauge fixing (\ref{canonical-fixing1})
together with the vanishing condition on the gauge field $h_M(\hat D)$ for the gauged dilatation in $G_c$,\footnote{The index $M$ covers the Lorentz vector and spinor indices $(m,\mu,\dot\mu)$.}
\begin{align}
& \bsh_M(\hat D)=0, \label{canonical-fixing2}
\end{align}
exhausts the $G_c$ degrees of freedom,\footnote{Note that the chirality of $\bsC,\bar\bsC$ is defined with respect to the superconformal covariant derivatives, which differ from the ones of Poincar\'e superspace, containing only the spin connection (i.e. the gauge field for Lorentz transformations). This is why the gauge fixing \eqref{canonical-fixing1} is allowed.} and 
reduces the conformal superspace to the so-called K\"ahler superspace  \cite{Binetruy:2000zx} characterised by $\cK$.\footnote{More explicitly, the gauge fixing \eqref{canonical-fixing1} along with \eqref{canonical-fixing2} converts the covariant derivatives of the conformal superspace to the ones of the K\"ahler superspace. 
The covariant derivatives in the conformal superspace contain the gauge fields for the Lorentz transformation, dilatation, chiral ${\rm U}(1)$ rotation, and special conformal transformations, while those in the K\"ahler superspace characterised by $\cK$ contain the gauge fields for the Lorentz transformations and the K\"ahler connection determined by $\cK$. The K\"ahler connection originates from the gauge field of the chiral ${\rm U}(1)$ rotation.}
In this superspace, a general (Poincar\'e) supergravity action of matter chiral superfields $\bsPhi$, with canonically normalised kinetic terms of the gravity multiplet, is written in terms of $\cK$ and a superpotential $\cW$ as  \cite{Binetruy:2000zx}
\begin{align}
-3\ka^{-2} \int_{\cK} \! d^4xd^4\te \, \bsE + \Big(
\ka^{-3} \int_{\cK} \! d^4xd^4\te \, \frac{\bsE}{\bsR} e^{\ka^2 \cK(\bsPhi,\bar\bsPhi)/2}\cW(\bsPhi)
+ {\rm h.c.} \Big),
\label{action-Kahler}
\end{align}
where `$\cK$' in the integral symbol indicates the K\"ahler superspace characterised by $\cK$.\footnote{The superfield $\bsR$ comes from the gauge-fixed special conformal  gauge superfield \eqref{hK-R}.} 
A complete component action of \eqref{action-Kahler} is given in \cite{Binetruy:2000zx} and is the same as the corresponding component action in Wess and Bagger \cite{Wess:1992cp}. In particular, 
the F-term scalar potential is given by the standard formula
\begin{align}
\kappa^4 \cV_F = e^{\ka^2\cK}(g^{\Phi\bar\Phi}D_\Phi \cW D_{\bar\Phi}\ol \cW - 3\ka^2\cW\ol \cW)|,
\label{scpot-sugra}
\end{align}
where $D_\Phi \cW=\pd_\Phi \cW+\ka^2(\pd_\Phi \cK)\cW$ and $g^{\Phi\bar\Phi}=(\pd_\Phi\pd_{\bar\Phi}\cK)^{-1}$.
Appendix~\ref{app:conf-sugra} contains a brief summary of conformal supergravity in conformal superspace.

\bigskip

\subsubsection*{Strategy}
Combining these facts, we may summarise the outline of our computation as follows:
\begin{enumerate}
\item Write down a UV action with ${\rm U}(1)_{\rm m}$ invariance in conformal superspace,
\begin{align}
S[\bsC; \bsPhi_+,\bsPhi_-,\bsV],
\end{align}
where $\bsC$ is the chiral compensator.
Note that setting $\bsC=\ol\bsC=1$ must recover the action \eqref{matterYM-sugra}.
The ${\rm U}(1)_{\rm m}$ invariance is as easy to implement as in the global supersymmetry case.

\item Adopt the unitary gauge $\bsPhi_-=v$ to fix the ${\rm U}(1)_m$ degrees of freedom,\footnote{This does not affect the superconformal invariance because $\bsPhi_-$ is taken to be superconformally invariant, as we will see shortly.} and integrate out the heavy fields to find an effective action,
\begin{align}
e^{-S_{\rm eff}[\bsC;\bsPhi_+]} = \int [d\bsV] \, e^{-S[\bsC;\bsPhi_+,v,\bsV]}.
\end{align}
The effective action $S_{\rm eff}$ is superconformal invariant assuming an invariant measure $[d\bsV]$.
Therefore, $S_{\rm eff}$ is still a conformal supergravity action.

\item Find $\cK_{\rm eff}$ for which the gauge fixing $\bsC=\ol\bsC=e^{\ka^2\cK_{\rm eff}/6}$ results in the action 
\begin{align}
-3\ka^{-2}\int_{\cK_{\rm eff}} \! d^4xd^4\te \, \bsE + 
\Big( \ka^{-3}\int_{\cK_{\rm eff}} \! d^4xd^4\te \, \frac{\bsE}{\bsR}e^{\cK_{\rm eff}/2}\cW_{\rm eff} + {\rm h.c.} \Big).
\label{kahler-action}
\end{align}
Note that the integrals are over the K\"ahler superspace characterised by $\cK_{\rm eff}$.

\item Compute the (F-term) scalar potential with the formula
\begin{align}
\kappa^4 \cV_F = e^{\ka^2\cK_{\rm eff}}(g^{\Phi\bar\Phi}D_\Phi \cW_{\rm eff}D_{\bar\Phi}\ol \cW_{\rm eff} - 3\ka^2\cW_{\rm eff}\ol \cW_{\rm eff})|,
\label{scpot-eff}
\end{align}
where $D_\Phi \cW_{\rm eff}=\pd_\Phi \cW_{\rm eff}+\ka^2(\pd_\Phi \cK_{\rm eff})\cW_{\rm eff}$ and $g^{\Phi\bar\Phi}=(\pd_\Phi\pd_{\bar\Phi}\cK_{\rm eff})^{-1}$.
\end{enumerate}

\subsection{Theory with gauged ${\rm U}(1)_{\rm m}$ invariance}
The UV action in conformal superspace which becomes the action \eqref{matterYM-sugra} after fixing the conformal compensators as $\bsC=\ol\bsC=1$ is actually very easy to write down,
\begin{align}
S &= 
\frac{1}{4}\int \! d^4xd^2\te ~ \boldsymbol{\cE} \cF(\bsPhi_-)\bsW^\al\bsW_\al + {\rm h.c.}
\nn\\
&\quad
+ \kappa^{-3} m \int \! d^4xd^2\te ~ \boldsymbol{\cE}\bsC^3\bsPhi_+\bsPhi_- + {\rm h.c.}
\nn\\
&\quad
- 3\kappa^{-2} \int \! d^4xd^4\te \, 
\bsE\bsC\ol\bsC e^{-\kappa^2\cK_0/3 - (q_+-q_-)\bsV/3},
\label{matterYM-CC}
\end{align}
and takes exactly the same form as in the case with the super-Weyl compensators \cite{Kaplunovsky:1994fg}.

To explain the notation, we need to introduce two important classes of superfields in conformal superspace: chiral and primary. A chiral superfield $\bsPhi$  is defined with respect to the superconformally covariant spinor derivative $\bar\nab_\dal$ by 
\begin{align}
\bar\nab_\dal\bsPhi=0. 
\end{align}
A primary superfield $\bsPhi$ of charges $(\de,w)$ is defined by
\begin{align}
\hat D\bsPhi=\de\bsPhi, \qquad \hat A\bsPhi=iw\Phi, \qquad \hat K_A\bsPhi=0,
\end{align}
where $\hat D,\hat A,\hat K_A$ are the generators for the dilatation, chiral ${\rm U}(1)$ rotation, and special conformal transformations.\footnote{The local Lorentz index $A$ in $\hat K_A$ stands for the vector and the undotted and dotted spinor indices $(a,\al,\dal)$. Therefore $\hat K_A$ denotes the generators $(\hat K_a,\hat S_\al,\hat{\bar{S}}^\dal)$.}

\medskip

We now explain the notation. For details, see Appendix~\ref{app:conf-sugra} and \cite{Butter:2009cp,Kugo:2016zzf,Kugo:2016lum}.
An action integral with $\int\!d^4xd^4\theta$ like the third line of \eqref{matterYM-CC} is called the D-type action.
Its integrand is required to be real primary of charge $(0,0)$ for gauge invariance. On the other hand, an action integral with $\int\!d^4xd^2\theta$ like the first and second lines of \eqref{matterYM-CC} is called the F-type action.
Its integrand is required to be chiral primary of charge $(0,0)$ for gauge invariance.

\medskip

The determinant $\bsE$ of the vierbein superfield is real primary of charges $(-2,0)$, while the determinant $\boldsymbol{\cE}$ of the ``chiral'' part of the vierbein superfield, called the chiral density, is chiral primary of charges $(-3,-2)$.

\medskip

The chiral superfields $\bsPhi_\pm$ are primary of charges $(0,0)$, transforming under the matter ${\rm U}(1)_{\rm m}$ as 
$\bsPhi_\pm \mapsto e^{\mp iq_\pm\bsLm}\bsPhi_\pm$, where $\bsLm$ is chiral primary of charges $(0,0)$.
The vector superfield $\bsV$ is primary of charges $(0,0)$, which transforms under ${\rm U}(1)_{\rm m}$ as
$\bsV \mapsto \bsV+i(\bsLm-\ol\bsLm)$.

The compensators $\bsC,\ol\bsC$ are chiral primary of charges $(1,2/3)$, and anti-chiral primary of charges $(1,-2/3)$, respectively. To guarantee ${\rm U}(1)_{\rm m}$ invariance, we need to assign ${\rm U}(1)_{\rm m}$ charges to the compensators $\bsC,\ol\bsC$ as
\begin{align}
\bsC \mapsto e^{i(q_+-q_-)\bsLm/3}\bsC, \qquad \ol\bsC \mapsto e^{-i(q_+-q_-)\ol\bsLm/3}\ol\bsC.
\label{U(1)onC}
\end{align}
$\cK_0$ is the gauge-invariant K\"ahler potential,
\begin{align}
\kappa^2\cK_0 = \ol\bsPhi_+ e^{q_+\bsV} \bsPhi_+ + \ol\bsPhi_- e^{-q_-\bsV} \bsPhi_-,
\end{align}
and $\bsW_\al$ is the chiral primary gaugino superfield of charges $(3/2,1)$, defined here with the superconformally covariant derivatives $\nab_\al,\bar\nab_\dal$ as\footnote{Note that $\nab_\al$ has charges $(1/2,-1)$ and $\bar\nab_\dal$ has $(1/2,1)$.}
\begin{align}
\bsW_\al = -\frac{1}{4}\bar\nab^2\nab_\al \bsV.
\end{align}

\subsection{Integrating out heavy fields}
We proceed to integrating out the heavy degrees of freedom.
For this, we first fix the matter ${\rm U}(1)_{\rm m}$ degrees of freedom by the unitary gauge $\bsPhi_-=v$, in which the action reads
\begin{align}
S &= 
\frac{1}{4}\int \! d^4xd^2\te ~ \boldsymbol{\cE}\bsW^\al \bsW_\al + {\rm h.c.}
\nn\\
&\quad
+ \kappa^{-3} mv \int \! d^4xd^2\te ~ \boldsymbol{\cE}\bsC^3\bsPhi_+ + {\rm h.c.}
\nn\\
&\quad
- 3\kappa^{-2} \int \! d^4xd^4\te \, 
\bsE\bsC\ol\bsC e^{-\kappa^2\cK/3},
\label{FI-unitary}
\end{align}
where we rescaled $\bsV$ to absorb the factor $1+b\ln v$, and 
$\cK$ is the gauge-fixed K\"ahler potential with the FI contribution,
\begin{align}
\kappa^2\mathcal{K} = \ol\bsPhi_+ e^{xq_-\bsV} \bsPhi_+ + v^2 e^{-q_-\bsV} + (x-1)q_-\bsV,
\end{align}
and we recall $x=q_+/q_-$.

We integrate out $\bsV$ at tree level by solving the equation of motion of $\bsV$ around its vacuum, neglecting higher derivative contributions.
The equation of motion of $\bsV$ reads
\begin{align}
\label{eomV}
-\kappa^2\nab^\al \bsW_\al +\bsC\ol\bsC e^{-\kappa^2\mathcal{K}/3}q_-\left( x\ol\bsPhi_+ e^{xq_-\bsV} \bsPhi_+ - v^2e^{-q_-\bsV} + x-1 \right) = 0.
\end{align}
As in the globally supersymmetric case in the last section, this equation of motion contains a tadpole. 
To integrate out $\bsV$ around its vacuum,  we first shift $\nab^\al \bsW_\al|$ to remove the tadpole, and then neglect the derivative term.
This gives the following low-energy effective equation of motion
\begin{align}
\label{eomnohatV}
\bsC\ol\bsC e^{-\kappa^2\mathcal{K}/3}q_-\left( x\ol\bsPhi_+ e^{xq_-\bsV} \bsPhi_+ - v^2e^{-q_-\bsV} + x-1 \right) -{q_-}\Delta  \simeq 0 .
\end{align}
Recall that $\De=x-1-v^2$.

We now integrate out $\bsV$ in the following way: It is convenient to rewrite the $\bsW\bsW$-part of \eqref{FI-unitary} using the formula \eqref{DtoF},\footnote{Its proof is outlined at the end of Appendix~\ref{app:conf-sugra}.}
\begin{align}
\frac{1}{4}\int \! d^4xd^2\te ~ \boldsymbol{\cE}\bsW^\al \bsW_\al+{\rm h.c.} = -\frac{1}{2}\int d^4xd^4\te \, \bsE\bsV\nab^\al \bsW_\al,
\label{WWtoWDV}
\end{align}
and then eliminate $\nab^\al \bsW_\al$ by substituting the exact equation of motion \eqref{eomV}. The first and third terms of the action \eqref{FI-unitary} then become
\begin{align}
\label{actionnoDDDD}
&\quad
\int \! d^4xd^4\te \, \bsE\left(
-\frac{1}{2}\bsV\nab^\al \bsW_\al -3\kappa^{-2}\bsC\ol\bsC e^{-\kappa^2\mathcal{K}/3}
\right)
\nn\\
&=
\kappa^{-2}\int \! d^4xd^4\te \, \bsE\bsC\ol\bsC e^{-\kappa^2\mathcal{K}/3} \nn\\
&\qquad\qquad
\times\Big[
-\frac{1}{2} q_-\bsV\left(x\ol\bsPhi_+ e^{xq_-\bsV} \bsPhi_+ - v^2e^{-q_-\bsV} + x-1\right)
 -3
\Big].
\end{align}
Next, combining the (low-energy) equation of motion \eqref{eomnohatV} with the second line of \eqref{actionnoDDDD}, we obtain the low-energy effective action,
\begin{align}
S_{\rm eff}[\bsC;\bsPhi_+]&= \kappa^{-3} mv \int \! d^4xd^2\te ~ \boldsymbol{\cE}\bsC^3\bsPhi_+ + {\rm h.c.} \nn\\
&\qquad  + \kappa^{-2}
\int \! d^4xd^4\te \, \bsE\Big(-\frac{1}{2}\Delta q_-\bsV - 3\bsC\ol\bsC e^{-\kappa^2\mathcal{K}/3}\Big),
\label{effaction}
\end{align}
where $\bsV$ must be understood to be a function of $\bsPhi_+$, determined by the equation of motion \eqref{eomnohatV}.

\subsection{Effective K\"ahler potential and superpotential}
Let us now fix the compensators.
As outlined at the end of Section~\ref{subsec:confsugra}, we find $\cK_{\rm eff}$ such that the gauge fixing 
\begin{align}
\label{CCfix}
\bsC=\ol\bsC=e^{\ka^2\cK_{\rm eff}/6}
\end{align}
makes the effective action \eqref{effaction} into the one of \eqref{kahler-action} in the K\"ahler superspace characterised by $\cK_{\rm eff}$, from which the scalar potential is given by the standard formula \eqref{scpot-eff}.
It is easy to see that this is realised by\footnote{One might have wondered why the equations \eqref{eomnohatV}, \eqref{actionnoDDDD} and \eqref{effaction} have terms proportional to $\bsV$ with weights $(0,0)$ despite the condition that they should have weight $(2,0)$ for the action to be superconformally invariant. This is because we have taken a heuristic route to find the effective K\"ahler potential and superpotential, keeping the compensators $\bsC,\ol\bsC$ undetermined, while we  used the D-tadpole subtraction in \eqref{eomnohatV}. An unambiguous way would be to fix the compensators as \eqref{CCfix} with \eqref{Keff_sugra} from the beginning, and follow the same steps as in the last subsection with fixed  compensators in the K\"ahler superspace characterised by $\cK_{\rm eff}$. This leads to the effective action \eqref{kahler-action} with the effective K\"ahler potential and superpotential \eqref{Keff_sugra}. It would be interesting to find another way of removing the tadpole that keeps the superconformal covariance.}
\begin{align}
\kappa^2\cK_{\rm eff} = \kappa^2 \cK + 3\ln\Big(1-\frac{1}{6} \Delta q_-\bsV\Big), \quad
\kappa^3\cW_{\rm eff} = mv\bsPhi_+,
\label{Keff_sugra}
\end{align}
where we used the formula \eqref{FtoD} which converts an F-type integral to a D-type one, and the identity on the gauge fixing of the chiral projection operator \eqref{chiralproj}.
The second term in the effective K\"ahler potential is the supergravity modification to the corresponding equation in the case of global supersymmetry \eqref{globalactionnoDDDD}, obtained in the limit $|\Delta| \ll 1$.

Indeed, a globally sypersymmetric limit is obtained in the limit $\kappa\to 0$, by defining the dimensionless supergravity parameters $v_{\rm sugra}$ and $\Delta_{\rm sugra}=x-1-v^2_{\rm sugra}$ in terms of the corresponding dimensionful parameters of the rigid theory $v_{\rm susy}$ and $\Delta_{\rm susy}=\xi-v_{\rm susy}^2$ as
\begin{align}
v_{\rm sugra}^2=\kappa^2v_{\rm susy}^2,\quad \quad
\Delta_{\rm sugra}=\kappa^2\Delta_{\rm susy}\,.
\label{rigidlimit}
\end{align}
The effective K\"ahler potential \eqref{Keff_sugra} and the extremisation condition \eqref{determine_q} then lead to the globally supersymmetric ones \eqref{Keff_global} and \eqref{global_const_eq}, respectively. Combining the two relations in \eqref{rigidlimit} gives $\xi=\kappa^{-2}(x-1)$. However, this implies in general that $\xi$ is not kept finite in the limit $\kappa \rightarrow 0$. 
The finiteness of $\xi$ can be reconciled only when we take the limit $x \rightarrow 1$ as $\kappa \rightarrow 0$. This implies that in the global limit the ${\rm U}(1)$ becomes an ordinary one (not gauged R-symmetry) and $\xi$ is arbitrary.

The gauge fixing \eqref{CCfix} simplifies the effective equation of motion \eqref{eomnohatV}  into
\begin{align}
\label{eff_eom}
\left( 1 - \frac{1}{6} \Delta q_-\bsV \right) \left( x\ol\bsPhi_+ e^{xq_-\bsV} \bsPhi_+ - v^2e^{-q_-\bsV} + x-1 \right)
- \Delta = 0,
\end{align}
which can be solved analytically for $\ol\bsPhi_+\bsPhi_+$ as a function of $\bsV$,
\begin{align}
\ol\bsPhi_+\bsPhi_+ &= x^{-1}e^{-x q_- \bsV} \Big( v^2 e^{-q_-\bsV} - x+1 + \frac{\Delta}{1- \frac{1}{6}\Delta q_- \bsV}\Big)\nonumber\\ 
&= x^{-1}e^{-x q_- \bsV} \Big( v^2 e^{-q_-\bsV} - v^2 + \frac{\frac{1}{6}\Delta^2q_- \bsV}{1- \frac{1}{6}\Delta q_- \bsV}\Big). 
\label{Phi_sol}
\end{align}
In the global limit $\ka\to 0$ with $x \rightarrow 1$, under the redefinition \eqref{rigidlimit} along with $\bsPhi_+^{\rm sugra}=\ka\bsPhi_+^{\rm susy}$, this solution is reduced to \eqref{globalSol_Phi} in terms of the dimensionful quantities of the rigid theory $v_{\rm susy}^2$ and $\bsPhi_+^{\rm susy}$.

Note that another non-trivial globally supersymmetric limit may be obtained by relaxing the first relation of \eqref{rigidlimit} and then  by matching \eqref{Delta_local} and \eqref{Delta_global} that fix $v_{\rm sugra}$ and $v_{\rm susy}$ as functions of the model parameters $(m_{\rm sugra},\kappa)$ in the local case and $(m_{\rm susy},\xi)$ in the global case, while the relation \eqref{Phi_sol} is reduced to \eqref{globalSol_Phi} by the second relation of \eqref{rigidlimit}.

\section{Inflation from the effective low-energy theory}
\label{sec:eff-inflation}

In Section~\ref{sec:FI-global}, we obtained the effective scalar potential of the FI model based on a gauged R-symmetry by integrating out the heavy degrees of freedom within global supersymmetry. However, as shown there,  the resulting model does not fit the class of inflation models discussed in Section~\ref{sec:SUSY-inflation} because the condition for the integration out cannot be reconciled with the condition that the scalar potential has a local maximum at the origin. In this section, we show that the model in the last section obtained by a similar procedure within supergravity does not have this problem and gives inflation models in the class discussed in Section~\ref{sec:SUSY-inflation}.

Strictly speaking, the effective theory found in the last section does not have a gauged R symmetry.
Therefore, to construct inflation models of the type we discussed in Section~\ref{sec:SUSY-inflation}, we need to add another gauged R symmetry to the low-energy theory, which we denote by ${\rm U}(1)^\prime$.  This can be achieved by extending the symmetry of the UV theory from ${\rm U}(1)_{\rm m}$ to ${\rm U}(1)_{\rm m}\times{\rm U}(1)^\prime$.  We assume that ${\rm U}(1)^\prime$ acts as a spectator during the integrating out process and survives as the gauged R-symmetry of the low-energy theory.  As summarised in Table \ref{charge}, $\bsPhi_+$ transforms under ${\rm U}(1)_m\times{\rm U}(1)^\prime$ with charge $(q_+,q)$ while $\bsPhi_-$ is singlet under ${\rm U}(1)^\prime$.

In what follows, we will analyse the behaviour of the effective K\"ahler potential around the origin and identify the parameter regions in which the scalar potential has a local maximum at the origin.

\begin{table}
\centering 
 \begin{tabular}{c|c|c}
  & ${\rm U(1)}_m$ & ${\rm U(1)}^\prime$   \\
\hline
$\bsPhi_+$ & $+q_+$ & $q$  \\
\hline
$\bsPhi_- $ & $-q_-$  & $0$   \\
\end{tabular}
\caption{\small The chiral multiplet $\bsPhi_+$ and $\bsPhi_-$ are charged under ${\rm U(1)}_m\times {\rm U(1)}^\prime$.  Note that ${\rm U(1)}^\prime$ does not play any role during the integrating out process and becomes R-symmetry of the low-energy theory.}
\label{charge}
\end{table}

\subsection{Perturbative analysis near the origin}

For simplicity, we absorb $q_-$ into the vector multiplet.\footnote{More precisely, we first rescale $q_-$ as $q_- \to q_-(1+b\ln v)^{-1/2}$ and then rescale $V$ as $V \to q_-V$ with the rescaled $q_-$ in the unitary gauge action \eqref{FI-unitary}.} To obtain the behaviour around the origin, we should first solve for $\bsV$ in terms of $\bar\bsPhi_+\bsPhi_+$ from equation (\ref{eff_eom}) perturbatively in the form
\begin{equation}
\label{pert_sol}
\bsV = V_0 + V_1\bar\bsPhi_+\bsPhi_+ + V_2 (\bar\bsPhi_+\bsPhi_+)^2 + V_3 (\bar\bsPhi_+\bsPhi_+)^3  + ... ~~.
\end{equation}
Substituting this into equation (\ref{eff_eom}) we obtain an explicit expression for the coefficients,
\begin{align}
V_0 =& \quad 0, \qquad V_1 = \frac{6 x}{\Delta^2-6 v^2} ,\nn\\ 
V_2 =&\quad \frac{6 x^2}{\left(\Delta^2-6 v^2\right)^3} \Big(-\Delta^3+6 \Delta^2 x-18 v^2 (2 x+1)\Big),\nn\\
V_3 =& \quad \frac{6 x^3 }{\left(\Delta^2-6 v^2\right)^5}\Big(\Delta^6-18 \Delta^5 x+6 \Delta^4 \left(v^2+9 x^2\right)+36 \Delta^3 v^2 (3 x+2) \nn\\
          & \quad -36 \Delta^2 v^2 (18x^2+9x-1)+216 v^4 (18x^2+9x+2)\Big).
\end{align}
Substituting the perturbative solution (\ref{pert_sol}) into the effective K\"ahler potential (\ref{Keff_sugra}), we obtain the effective K\"ahler potential around the local maximum,
\begin{align}
\label{Kpert1}
\kappa^2\mathcal{K}_{\rm eff} = v^2 + K_1\bar\bsPhi_+\bsPhi_+ + K_2 (\bar\bsPhi_+\bsPhi_+)^2  +  K_3 (\bar\bsPhi_+\bsPhi_+)^3 +... ~ ,
\end{align}
where the first three coefficients read
\begin{align}
K_1 =&~ \frac{\Delta^2+3 \Delta x-6 v^2}{\Delta^2-6 v^2} ,\\ 
K_2 =&~ -\frac{3 x^2 \left(-\Delta^4-12 \Delta^3 x+30 \Delta^2 v^2+36 \Delta v^2 (2 x+1)-72 v^4\right)}{2 \left(\Delta^2-6 v^2\right)^3}, \\
K_3 =&~ \frac{x^3 }{\left(\Delta^2-6 v^2\right)^5} \Big\{-\Delta^7-18 \Delta^6 x+6 \Delta^5 \left(8 v^2+27 x^2\right) -18 \Delta^4 v^2 (12 x-7) \nn\\
          &~ -36 \Delta^3 v^2 \left(v^2+54 x^2+27 x-3\right)+108 \Delta^2 v^4 (24 x+7) \nn\\
          &~ +648 \Delta v^4 \left(9 x^2+9 x+2\right)-1296 v^6 (3 x+1)\Big\}.
\end{align}
We then define the canonically normalized chiral superfield $\bsPhi$ as
\begin{equation}
\label{chiral}
\bsPhi :=  \sqrt{K_1}\;\bsPhi_+.
\end{equation} 
After absorbing the constant term  $v^2$ in (\ref{Kpert1}) by a K\"ahler transformation, the effective K\"ahler potential in $\bsPhi$ becomes
\begin{equation}
 \ka^2\mathcal{K}_{\rm eff} =  \ol{\bsPhi}\bsPhi +A_2 (\ol{\bsPhi}\bsPhi)^2 + A_3 (\ol{\bsPhi}\bsPhi)^3  + ... ~ ,
 \label{Keff_pert}
\end{equation}
where the first two nontrivial coefficients $A_2,A_3$ read
\begin{align}
A_2 =&~  \frac{3 x^2 \left(\Delta^4+12 \Delta^3 x-30 \Delta^2 v^2-36 \Delta v^2 (2 x+1)+72 v^4\right)}{2 \left(\Delta^2-6 v^2\right) \left(\Delta^2+3 \Delta x-6 v^2\right)^2}, \\
A_3 =&~ \frac{x^3}{\left(\Delta^2-6 v^2\right)^2 \left(\Delta^2+3 \Delta x-6 v^2\right)^3} \Big\{-\Delta^7-18 \Delta^6 x+6 \Delta^5 \left(8 v^2+27 x^2\right)\nn\\
         &~ +18 \Delta^4 v^2 (7-12 x)-36 \Delta^3 v^2 \left(v^2+54x^2+27x-3\right)\nn\\
         &~ +108 \Delta^2 v^4 (24 x+7)+648 \Delta v^4 (18x^2+9x+2)-1296 v^6 (3 x+1)\Big\}.
\end{align}

\begin{figure}
    \centering
    \begin{subfigure}[b]{0.46\textwidth}
        \includegraphics[width=\textwidth]{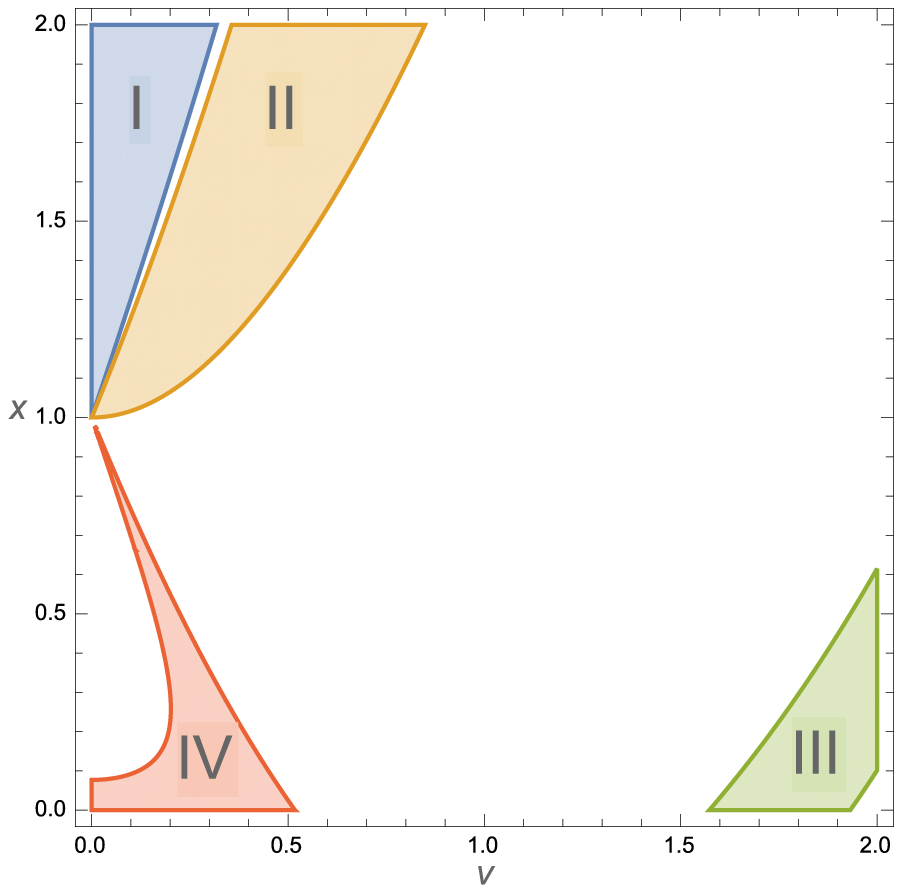}
        \caption{}
        \label{fig:a}
    \end{subfigure}
    ~ 
    \begin{subfigure}[b]{0.465\textwidth}
        \includegraphics[width=\textwidth]{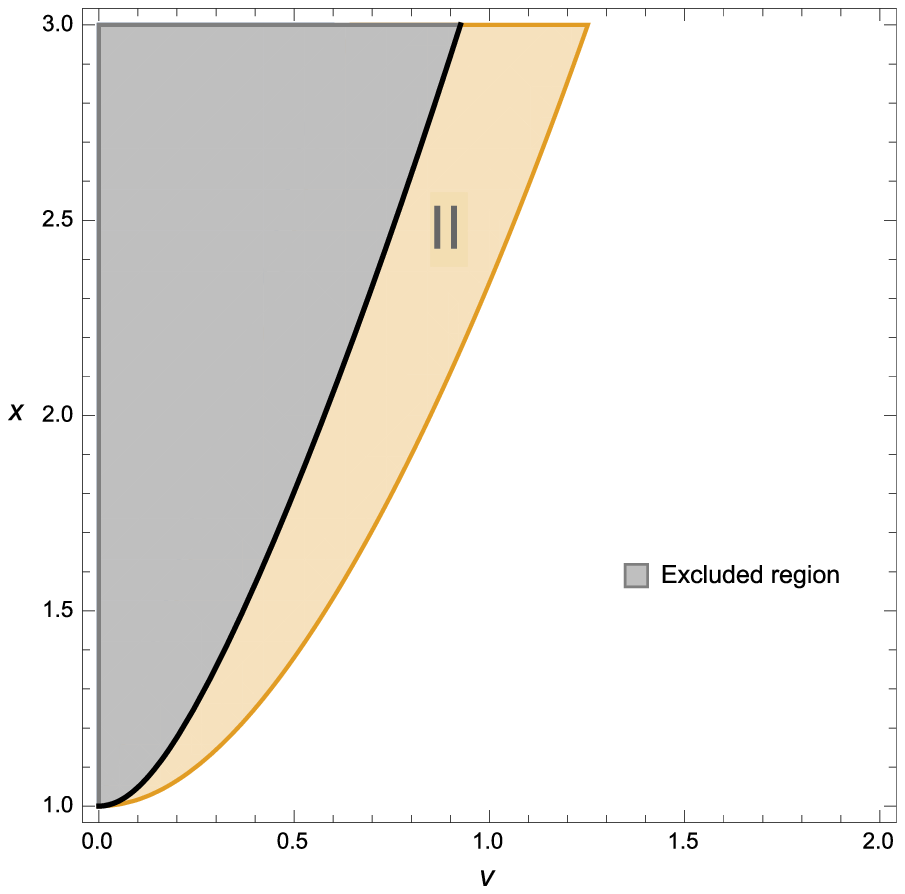}
        \caption{}
        \label{fig:b}
    \end{subfigure}
    \caption{(a) Allowed parameter space $(v,x)$ with $0 < v < 2.0$ and $0 < x < 2.0$.  The colored regions in which $A_2>0$ can be divided into 4 parts, namely I, II, III and IV.  (b) \textbf{Region I} and part of \textbf{Region II} are in the excluded area where $v^2 - \frac{1}{4} x (x-1-v^2) < 0$  where the integrating out condition is not  satisfied.}\label{plot01}
\end{figure}

The condition for having a local maximum at the origin is $A_2 > 0$.  In the two-dimensional parameter space $(v,x)$, the domain in which $A_2$ is positive can be divided into four regions according to the signs of $\Delta=x-1-v^2$ and of the scalar component $c = \bsV |$ in each region.  They are
\begin{itemize}
  \item \textbf{Region I}: with $\Delta > 0$, $c \geqslant 0$,
  \item \textbf{Region II}: with $\Delta >0$, $c \leqslant 0$,
  \item \textbf{Region III}: with $\Delta <0$, $c \leqslant 0$,
  \item \textbf{Region IV}: with $\Delta <0$, $c \geqslant 0$.
\end{itemize}
In Section \ref{sec:eff-potential}, we will show how the sign of $c$ is related to the reality condition on the inflaton.
These four regions are shown in Fig.~\ref{fig:a}.  In the next subsection, we will study the global minimum of the scalar potential for each region, and show that a Minkowski minimum is allowed in the presence of D-term in \textbf{Region I} and \textbf{III}, while \textbf{Region II} and \textbf{IV} have only de Sitter minimum with a large cosmological contant.  We will also show that the integrating out condition excludes \textbf{Region I}.  Therefore, this leaves \textbf{Region III} as the only possible domain for slow-roll inflation with a nearby minimum having a tuneable vacuum energy.

\subsection{The effective scalar potential and slow-roll parameters}
\label{sec:eff-potential}

In order to study the global minimum of the potential and compare our predictions for inflation with the observational data, we need the exact expression of the scalar potential.  Using the analytic solution (\ref{Phi_sol}) for $\ol\bsPhi{}_+\bsPhi_+$ as a function of $\bsV$, we will express the scalar potential as a function of  $c = \bsV |$ instead of $\varphi_+ = \bsPhi_+|$ .   

Combining the effective K\"ahler potential (\ref{Keff_sugra}) with the analytic solution \eqref{Phi_sol}, we express the effective K\"ahler potential as a function of the vector multiplet $\bsV$,
\begin{align}
\label{KeffV}
\kappa^2\mathcal{K}_{\rm eff}(\bsV) &= \frac{1}{x}\Big[v^2(1+x)e^{-\bsV}+\frac{\Delta}{1-\frac{1}{6}\Delta\bsV}-x+1 \Big]+(x-1)\bsV \nn\\
& \qquad + 3 \ln\Big[1-\frac{1}{6} \Delta\bsV \Big].
\end{align}
Note that $\bsV$ must be understood as a function of $\ol\bsPhi{}_+\bsPhi_+$ when we compute the scalar potential, using for instance eq.~\eqref{scpot-sugra}.
The effective superpotential is
\begin{equation}
\kappa ^3 \mathcal{W}_{\rm eff} = mv \bsPhi_+.
\label{E_eff}
\end{equation}
Using the formula \eqref{Dpot} and expressing it in the D-term potential in terms of $c=\bsV|$ instead of $\vphi_+=\bsPhi_+|$, we find the low energy D-term potential given by
\begin{align}
\kappa^4\mathcal{V}_D &= \frac{y^2e^{-2 c} m^2 v^2}{8 x^2 } \Big[\rho  v^2 (x+1-xe^c)  c' -2 e^c x \nn\\
& \qquad\qquad\qquad\quad
-e^c \rho  c' \frac{x\Delta (3-c \Delta)}{6-c\Delta}-\frac{6e^c \rho  c'\Delta^2}{(6-c\Delta)^2} \Big]^2,
\label{D-termV}
\end{align}
where we introduced a new parameter $y := q/mv$. Recall also that $\Delta = x-1-v^2$.  The new field variable $\rho$ is defined as $\rho := (\varphi^*_+\varphi_+)^{1/2}$, which stands for the inflaton. This can be written in terms of $c$ with the help of (\ref{Phi_sol}) as  
\begin{align}
\rho^2 = \frac{e^{-x c}}{x} \Big[ v^2 e^{-c} - x + 1 + \frac{\Delta}{1- \frac{1}{6} \Delta c}\Big].
\label{rho_square}
\end{align}
For any given value of the parameters $v$ and $x$, we can choose the ``physical domain" of $c$ in such a way that $\rho^2 > 0$.
We also introduced $c' = d c/d \rho$, $c'' = d^2 c/d\rho^2$, which can be expressed in terms of $c$ with the help of (\ref{rho_square}) as
\begin{align}
c' &= \frac{2 \rho  x (6-c \Delta) e^{c (x+1)}}{e^c \Delta ^2-v^2 \left(6-\left(c+e^c-1\right) \Delta\right)-\rho ^2 x e^{c (x+1)} (6x -c \Delta  x-\Delta)},\\
c'' &= -\frac{v^2 (6-c\Delta+2\Delta) \left(c'\right)^2}{e^c\Delta ^2-v^2 \left(6-\left(c+e^c-1\right) \Delta\right)- \rho ^2 x e^{c (x+1)} (6x-c \Delta x-\Delta )}\nn\\
   &+\frac{x e^{c (x+1)} \left(\rho  c' \left(\rho  x c' (x (6-c \Delta)-2 \Delta )+4 (x (6-c \Delta)-\Delta )\right)+2 (6-c \Delta)\right)}{e^c\Delta ^2-v^2 \left(6-\left(c+e^c-1\right) \Delta\right)- \rho ^2 x e^{c (x+1)} (6x-c \Delta x-\Delta )}.
\end{align}
On the other hand, the effective F-term potential is given by
\begin{align}
\kappa^4\mathcal{V}_F &= m^2v^2 e^{\kappa^2 \mathcal{K}_{\rm eff}(c)}\Big( -3\rho^2 + \frac{4\mathcal{A}^2(c)}{\mathcal{B}(c)} \Big),
\label{F-termV}
\end{align}
where we introduced two functions $\cA(c),\cB(c)$
\begin{align}
\mathcal{A}(c) =&~ 1+\frac{\rho  c'}{2 x (6-c \Delta)^2}e^{-c}\Big[6 e^c v^4+e^c \xi  \left(x \left(c^2 \Delta ^2-9 c \Delta +18\right)+6 \xi \right) \label{Ac} \nn\\
             &~+v^2 \left(-c^2 \Delta ^2+12 c \Delta -12 e^c \xi +x (6-c \Delta) \left(c \Delta +3 e^c-6\right)-36\right)\Big],\\
\mathcal{B}(c) =&~ -\frac{3 \Delta  \left(\rho  c''+c'\right)}{\rho  (6-c \Delta)}+\frac{\xi  \left(\rho  c''+c'\right)}{\rho }+\frac{\left(\rho  c''+c'\right) \left(\frac{6 \Delta ^2}{(6-c \Delta)^2}-e^{-c} v^2 (x+1)\right)}{x\rho }\nn\\
&~ + \frac{\left(c'\right)^2}{x} \left(e^{-c} v^2 (x+1)+\frac{12 \Delta ^3}{(6-c \Delta)^3}\right) -\frac{3 \Delta ^2 \left(c'\right)^2}{(6-c \Delta)^2}. \label{Bc}
\end{align}

\medskip

To compute the slow-roll parameters, we need the canonically normalised inflaton field $\chi$ defined through $\chi^\prime := \frac{d\chi}{d\rho}=\sqrt{2 g_{\bar\bsPhi_+ \bsPhi_+}}$, which can be written in terms of $c$ as
\begin{align}
\chi^\prime = \kappa\sqrt{\left(\frac{c^\prime}{2\rho} + \frac{c^{\prime\prime}}{2}\right) \frac{d}{dc}\mathcal K_{\rm eff}(c)  +\frac{(c^\prime)^2}{2\rho} \frac{d^2}{dc^2}\mathcal K_{\rm eff}(c) }.
\end{align}
The slow-roll parameters $\epsilon$ and $\eta$ are given in terms of $c$ by
\begin{align}
\epsilon =&~ \frac{1}{2\kappa^2}\left(\frac{d\mathcal{V}/d\chi}{\mathcal V} \right)^2 =\frac{1}{2\kappa^2}\left( \frac{d \mathcal V/dc}{\mathcal V} \frac{c^\prime}{\chi^\prime}\right)^2, \\
\eta =&~ \frac{1}{\kappa^2}\frac{d^2\mathcal{V}/d\chi^2}{\mathcal V}, \nonumber\\
      =&~ \frac{1}{\kappa^2}\left( \frac{d^2 \mathcal V/dc^2}{\mathcal V}\left(\frac{c^\prime}{\chi^\prime}\right)^{2} + \frac{d \mathcal V/dc}{\mathcal V} \frac{c^{\prime\prime}}{\chi^\prime} - \frac{d\mathcal V/dc}{\mathcal V}\frac{d\chi^\prime/dc}{\chi^\prime}  \left(\frac{c^\prime}{\chi^\prime}\right)^2\right).
\end{align}
The number of e-folds $N$ during inflation can be expressed as
\begin{align}
N = \int_{\chi_*}^{\chi_{\rm end}} \frac{\mathcal V}{\partial_{\chi} \mathcal{V}} d \chi =\int_{\rho_*}^{\rho_{\rm  end}} \frac{\mathcal V}{\partial_{\rho} \mathcal{V}} (\chi^\prime)^2 d \rho = \int^{c_*}_{c_{\rm end}} \frac{\mathcal V}{\partial_{c} \mathcal{V}}\left(\frac{\chi^\prime}{c^\prime} \right)^2 d c,
\end{align}
where we choose $|\eta(c_{\rm end})| = 1$ and $c_*$ is  the value of $c$ at the horizon exit.  Now we can compare the theoretical predictions of our model to the observational data,  specifically the power spectrum of primordial perturbations of the CMB, namely the amplitude of density fluctuations $A_s$, the spectral index $n_s$ and the tensor-to-scalar ratio of primordial fluctuations $r$. They can be written in terms of the slow-roll parameters:
\begin{align}
A_s =&\quad \frac{\kappa^4 \mathcal{V}_*}{24 \pi^2 \epsilon_*},\\
n_s =&\quad 1+2\eta_* - 6\epsilon_* \simeq 1+ 2\eta_*,\\
r =&\quad 16 \epsilon_*,
\end{align}
evaluated using the field value $c_*$ at the horizon exit.

\subsubsection{Region I}
\begin{figure}[h]
    \centering
    \begin{subfigure}[b]{0.46\textwidth}
        \includegraphics[width=\textwidth]{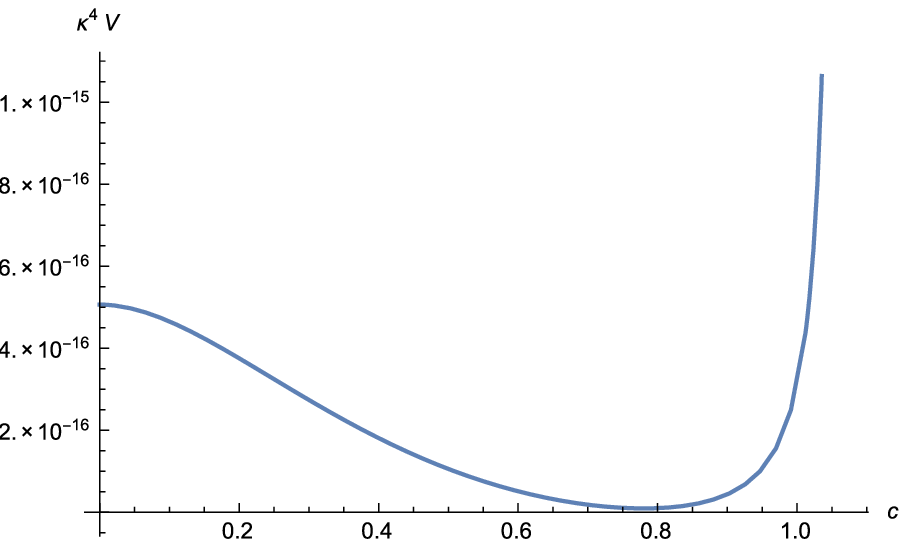}
        \caption{}
        \label{fig:Ia}
    \end{subfigure}
    ~ 
    \begin{subfigure}[b]{0.44\textwidth}
        \includegraphics[width=\textwidth]{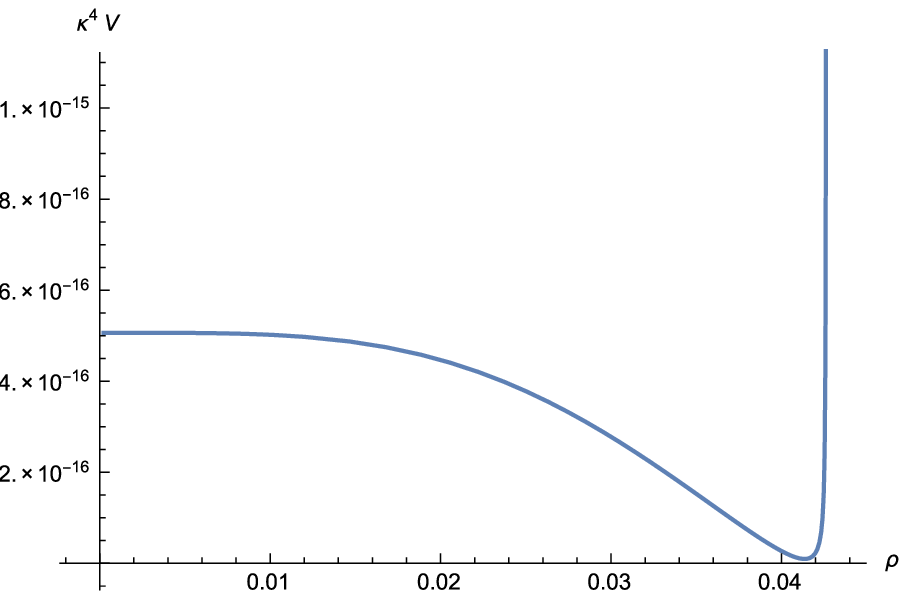}
        \caption{}
        \label{fig:Ib}
    \end{subfigure}
    \\ 
    \begin{subfigure}[b]{0.44\textwidth}
        \includegraphics[width=\textwidth]{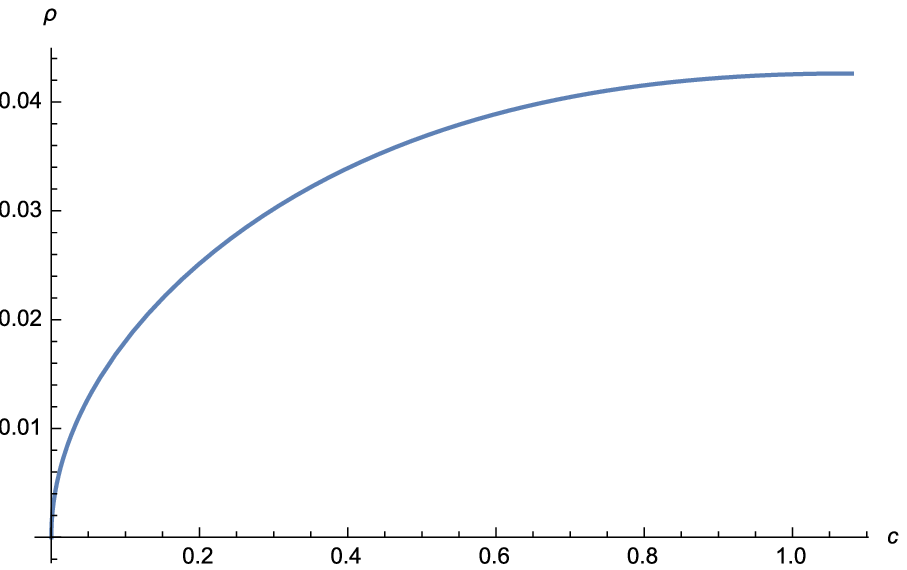}
        \caption{}
        \label{fig:Ic}
    \end{subfigure}
    ~ 
    \begin{subfigure}[b]{0.46\textwidth}
        \includegraphics[width=\textwidth]{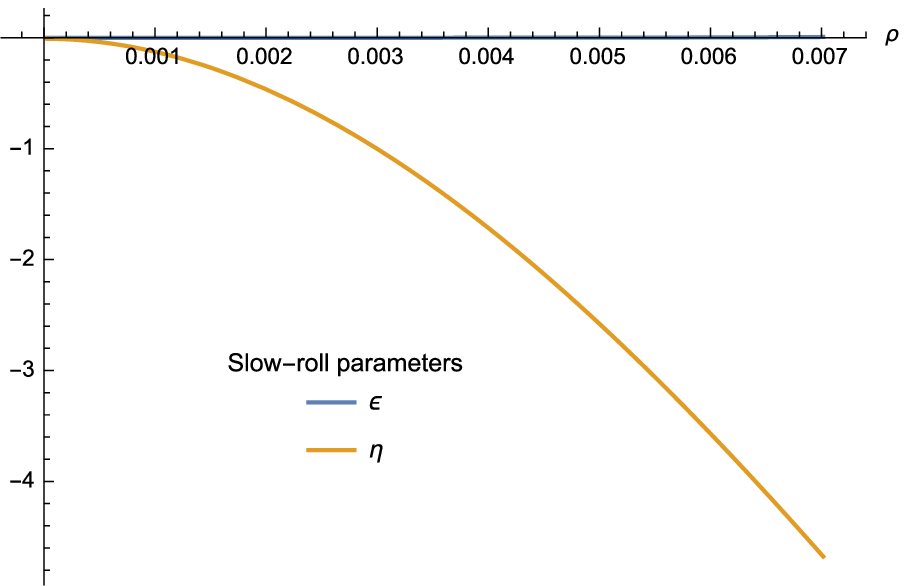}
        \caption{}
        \label{fig:Id}
    \end{subfigure}
    \caption{The scalar potential for a model in \textbf{Region I} of Fig.~\ref{plot01} with parameters (\ref{parameter:I}) is plotted as a function of $c$ and $\rho$ coordinates in (a) and (b) respectively.  Plot in (c) shows the relation between $\rho$ and $c$.  The slow-roll parameters $\epsilon$ and $\eta$ are shown in (d).}\label{fig:RegionI}
\end{figure}

We can choose for example
\begin{align}
v = 0.0999, \quad x = 1.3024, \quad y = 0.0769, \quad m = 1.4214 \times10^{-6}.
\label{parameter:I}
\end{align}
For this choice of parameters, we have $\Delta = 0.2924$.  Note that $m$ determines the overall scale of the scalar potential and is fixed using the amplitude $A_s$ from CMB data. From (\ref{determine_q}), we obtain $q_- \approx 5.3097 \times 10^{-6}$.  The scalar potential for these parameters as a function of $c$ or $\rho$ is plotted in Fig.~\ref{fig:Ia} and \ref{fig:Ib}, respectively.  The relation between $c$ and $\rho$ coordinates is shown in Fig.~\ref{fig:Ic}, from which we can see that the physical domain which guarantees the positivity of $\rho$ is $c > 0$. We plot the slow-roll parameter in $\rho$ coordinates in Fig.~\ref{fig:Id}.

Choosing the initial condition $c_* = 3.53\times 10^{-5}$ and $c_{\rm end} = 3.00\times 10^{-3}$ (or equivalently, by using (\ref{rho_square}), $\rho_* = 3.40\times 10^{-3}$ and $\rho_{\rm end} = 3.14\times 10^{-3}$), we obtain $N = 59.82$, $n_s = 0.9548$, $r = 1.53\times10^{-8}$ and $A_s = 2.2\times10^{-9}$, which are within the $2\sigma$-region of Planck'18 data \cite{Akrami:2018odb}.  

Let us now examine the particle mass spectrum in the UV theory for the parameter set in \textbf{Region I}.   By observing that $m,q_- \ll v,x$, we can show  using (\ref{mass_phi_p}) that the mass-square difference  between the vector field and $\varphi_+$ is
\begin{align}
m^2_{A_\mu} - m^2_{\varphi_+} \simeq q_-^2\left(v^2 - \frac{1}{4}x\frac{\Delta}{1+b\ln v} \right).
\end{align}
Note that $b$ is of order $q_-^2 \ll 1$ and can be neglected.  The parameter set $(v,x)$ which satisfies
\begin{align}
v^2 - \frac{1}{4}x(x-1-v^2) < 0
\end{align}
gives $m^2_{A_\mu}<m^2_{\varphi_+}$ and must be excluded as it violates the integrating out condition.  

In Fig.~\ref{fig:b}, we plot the excluded region in the parameter space $(v,x)$.   We can see that \textbf{Region I} and some part of \textbf{Region II} are in the excluded region and do not satisfy the integrating out condition.  We can show quantitatively by using (\ref{mass_phi_p}), (\ref{mass_phi_m2}) and (\ref{mass_phi_m1}) that the  parameters (\ref{parameter:I}) give the mass ratios,
\begin{align}
\left. \frac{m^2_{A_\mu}}{\cV^{\rm {UV}}_{\vphi_+^*\vphi_+}}\right|_{\rm vac} \approx 0.0595, ~ 
\left. \frac{\cV^{\rm {UV}}_{\vphi_-^*\vphi_-}}{\cV^{\rm {UV}}_{\vphi_+^*\vphi_+}}\right|_{\rm vac}\approx 0.7944, ~ 
\left. \frac{\cV^{\rm {UV}}_{\vphi_-^*\vphi_-^*}}{\cV^{\rm {UV}}_{\vphi_+^*\vphi_+}}\right|_{\rm vac}\approx 0.0236.
\end{align}
In conclusion, although the parameter set in Region I leads to a scalar potential that allows slow-roll inflation and Minkowski vacua, the effective K\"ahler potential can not be obtained consistently from integrating out heavy fields that we discussed in the previous sections.

\subsubsection{Region II}
\begin{figure}[h]
    \centering
    \begin{subfigure}[b]{0.42\textwidth}
        \includegraphics[width=\textwidth]{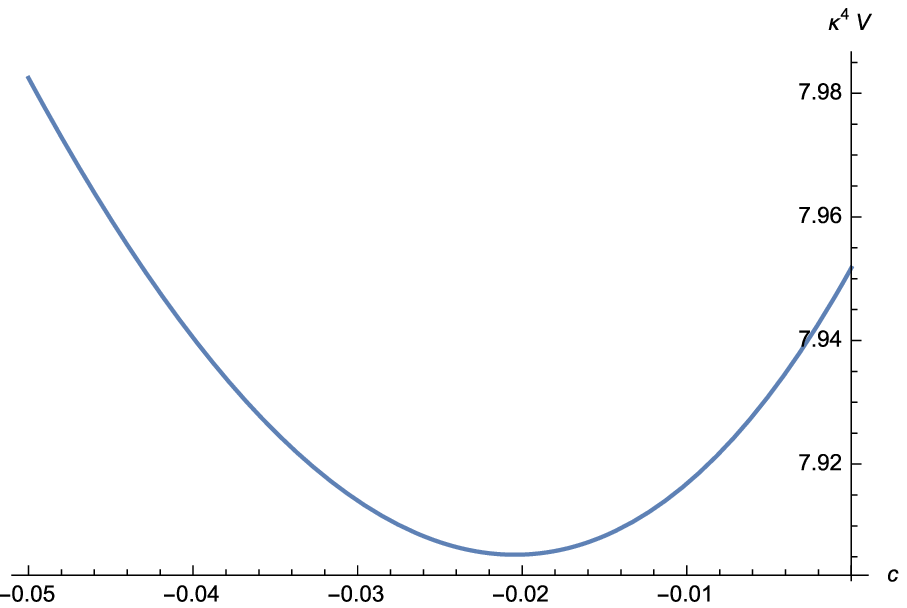}
        \caption{}
        \label{fig:IIa}
    \end{subfigure}
    ~ 
    \begin{subfigure}[b]{0.41\textwidth}
        \includegraphics[width=\textwidth]{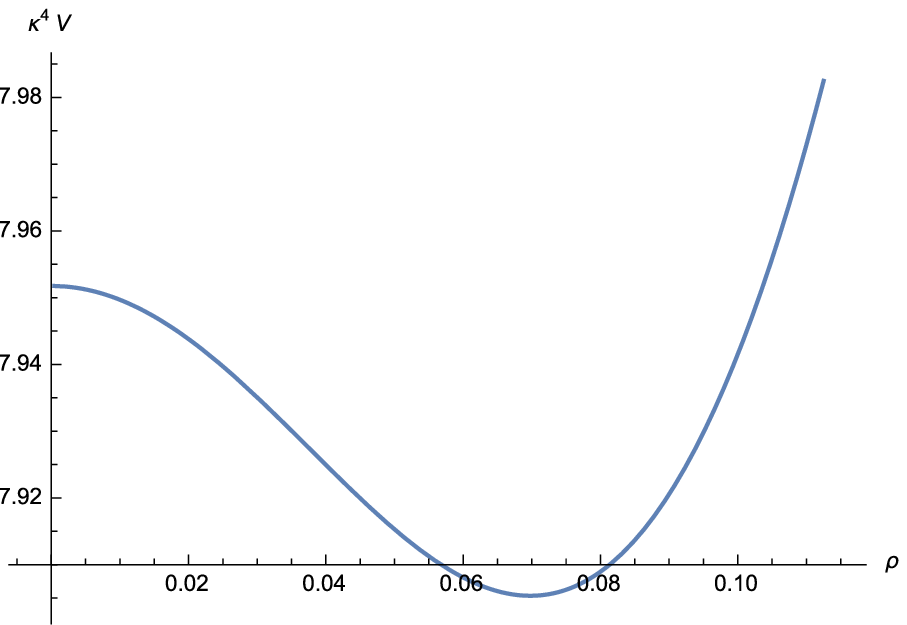}
        \caption{}
        \label{fig:IIb}
    \end{subfigure}
    \\ 
    \begin{subfigure}[b]{0.4\textwidth}
        \includegraphics[width=\textwidth]{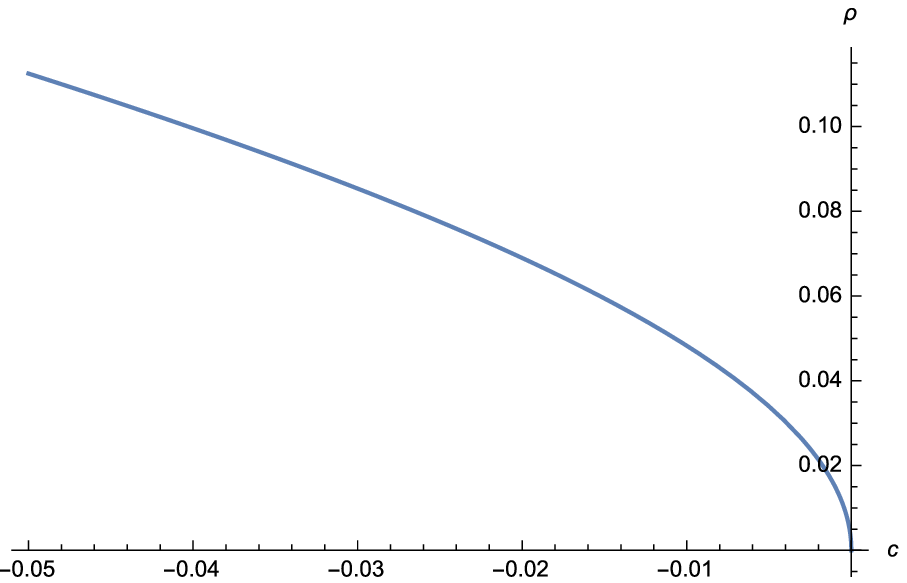}
        \caption{}
        \label{fig:IIc}
    \end{subfigure}
    \caption{The scalar potential for a model in \textbf{Region II} of Fig.~\ref{plot01} with parameters (\ref{parameter:II}) is plotted as a function of $c$ and $\rho$ coordinates in (a) and (b), respectively.  Plot in (c) shows the relation between $\rho$ and $c$. }\label{fig:RegionII}
\end{figure}

We choose parameters that are outside of the excluded region shown in Fig.~\ref{fig:b},  for example
\begin{align}
v = 0.60, \quad x = 1.55, \quad y = 0.0, \quad m = 3.00.
\label{parameter:II}
\end{align}
The scalar potential for these parameters as a function of $c$ and $\rho$ are plotted in Fig.~\ref{fig:IIa} and \ref{fig:IIb}, respectively.  The relation between $c$ and $\rho$ coordinates is shown in Fig.~\ref{fig:IIc} where in this case the physical domain is $c < 0$.  For this choice of parameters, we have $\Delta = 0.19$.  

Using (\ref{determine_q}), we obtain $q_- = 17.80$.  From (\ref{mass_phi_p}), (\ref{mass_phi_m2}) and (\ref{mass_phi_m1}), we find the mass ratios,
\begin{align}
\left.\frac{m^2_{A_\mu}}{\cV^{\rm {UV}}_{\vphi_+^*\vphi_+}}\right|_{\rm vac} \approx 2.9084, ~ \left. \frac{\cV^{\rm {UV}}_{\vphi_-^*\vphi_-}}{\cV^{\rm {UV}}_{\vphi_+^*\vphi_+}}\right|_{\rm vac}\approx 1.1410, ~ \left. \frac{\cV^{\rm {UV}}_{\vphi_-^*\vphi_-^*}}{\cV^{\rm {UV}}_{\vphi_+^*\vphi_+}}\right|_{\rm vac}\approx 1.5252.
\end{align}
Although we can find sets of parameters that satisfy the integrating out condition, the scalar potential does not allow for a global  minimum with small cosmological constant in this region.  

\begin{figure}[h]
    \centering
    \begin{subfigure}[b]{0.44\textwidth}
        \includegraphics[width=\textwidth]{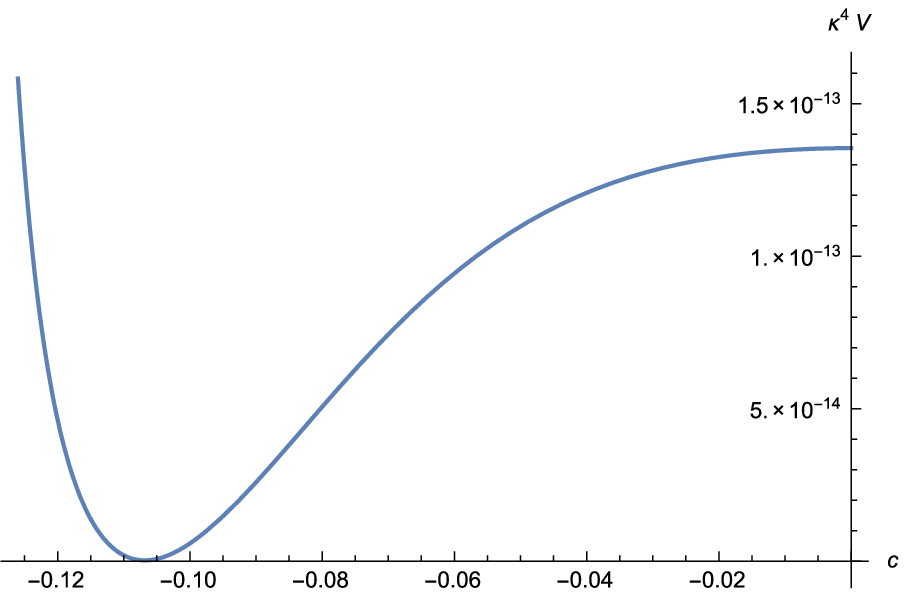}
        \caption{}
        \label{fig:IIIa}
    \end{subfigure}
    ~ 
    \begin{subfigure}[b]{0.44\textwidth}
        \includegraphics[width=\textwidth]{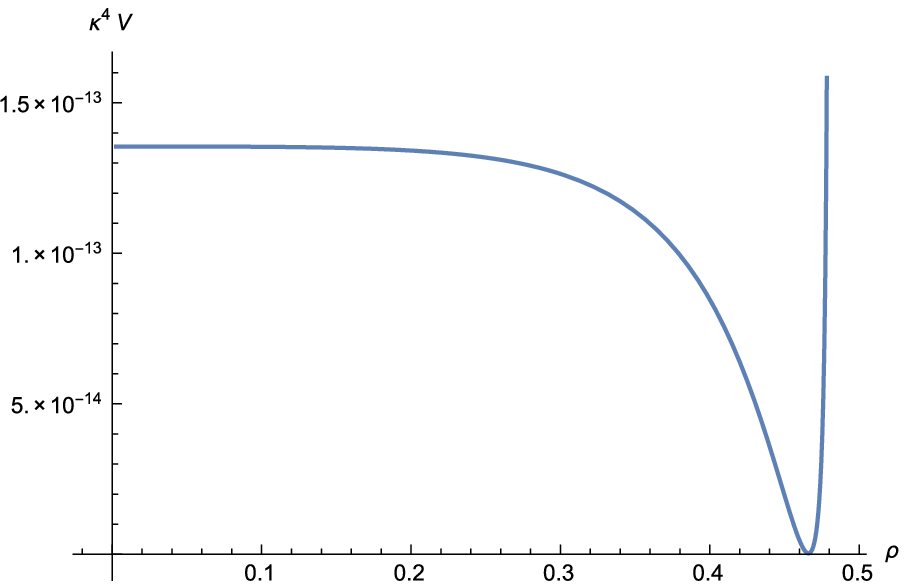}
        \caption{}
        \label{fig:IIIb}
    \end{subfigure}
    \\ 
    \begin{subfigure}[b]{0.44\textwidth}
        \includegraphics[width=\textwidth]{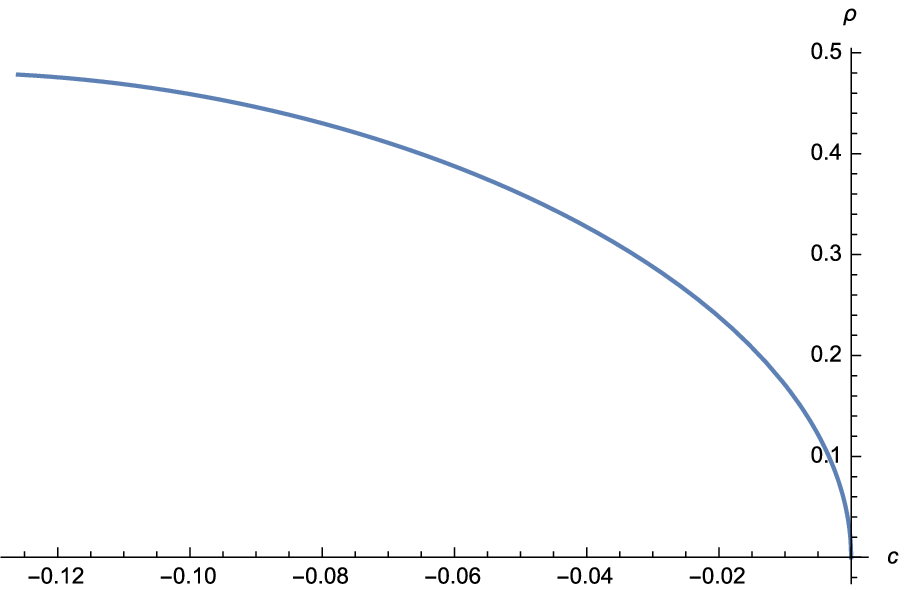}
        \caption{}
        \label{fig:IIIc}
    \end{subfigure}
    ~ 
    \begin{subfigure}[b]{0.46\textwidth}
        \includegraphics[width=\textwidth]{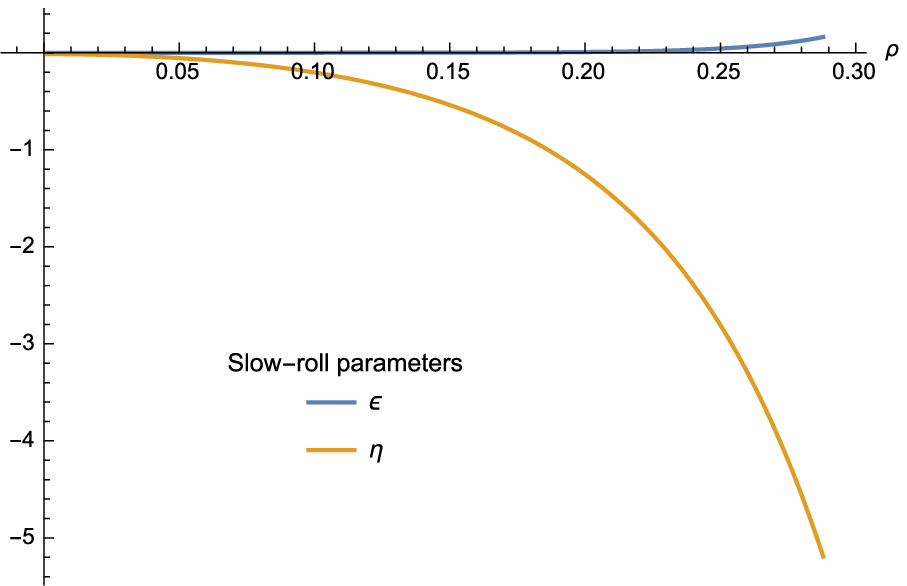}
        \caption{}
        \label{fig:IIId}
    \end{subfigure}
    \caption{Plots of the scalar potential 
    in \textbf{Region III} of Fig.~\ref{plot01} with parameters 
    (\ref{parameter:III}) as a function of the coordinate $c$ in (a) and $\rho$ in (b). The relation between $\rho$ and $c$ is plotted in (c).  The slow-roll parameters $\epsilon$ and $\eta$ are shown in (d).}\label{fig:RegionIII}
\end{figure}

  \subsubsection{Region III}
 
This case is not in the excluded region shown in Fig.~\ref{fig:b}, so the integration out condition may be satisfied. We can choose for example
\begin{align}
v = 1.86945, \quad x = 0.08435, \quad y = 4.07, \quad m = 3.77 \times10^{-8}.
\label{parameter:III}
\end{align}
For this choice, we have $\Delta = -4.41049$.

The scalar potential for these parameters as a function of $c$ and $\rho$ is plotted in Fig.~\ref{fig:IIIa} and \ref{fig:IIIb}, respectively.  The relation between $c$ and $\rho$ coordinates is shown in Fig.~\ref{fig:IIIc} where the physical domain is $c < 0$. The slow-roll parameters in $\rho$ coordinates are plotted in Fig.~\ref{fig:IIId}.

Choosing the initial condition $c_* = -0.00017$ and $c_{end} = -0.01192$ (or equivalently, by using (\ref{rho_square}), $\rho_* = 0.0225$ and $\rho_{end} = 0.1869$), we obtain $N = 59.48$, $n_s = 0.9597$, $r = 4.15\times10^{-6}$ and $A_s = 2.2\times10^{-9}$, which are within the $2\sigma$-region of Planck'18 data \cite{Akrami:2018odb}.

Using the constraint (\ref{determine_q}), we obtain $q_- \approx 31.5413$.  From (\ref{mass_phi_p}), (\ref{mass_phi_m2}) and (\ref{mass_phi_m1}), we find that the mass ratios indeed satisfy the integrating out condition,
\begin{align}
\left.\frac{m^2_{A_\mu}}{\cV^{\rm {UV}}_{\vphi_+^*\vphi_+}}\right|_{\rm vac} \approx 38.2253, ~ \left. \frac{\cV^{\rm {UV}}_{\vphi_-^*\vphi_-}}{\cV^{\rm {UV}}_{\vphi_+^*\vphi_+}}\right|_{\rm vac}\approx 21.9463, ~ \left. \frac{\cV^{\rm {UV}}_{\vphi_-^*\vphi_-^*}}{\cV^{\rm {UV}}_{\vphi_+^*\vphi_+}}\right|_{\rm vac}\approx 9.8853.
\end{align}

\subsubsection{Region IV}

\begin{figure}
    \centering
    \begin{subfigure}[b]{0.4\textwidth}
        \includegraphics[width=\textwidth]{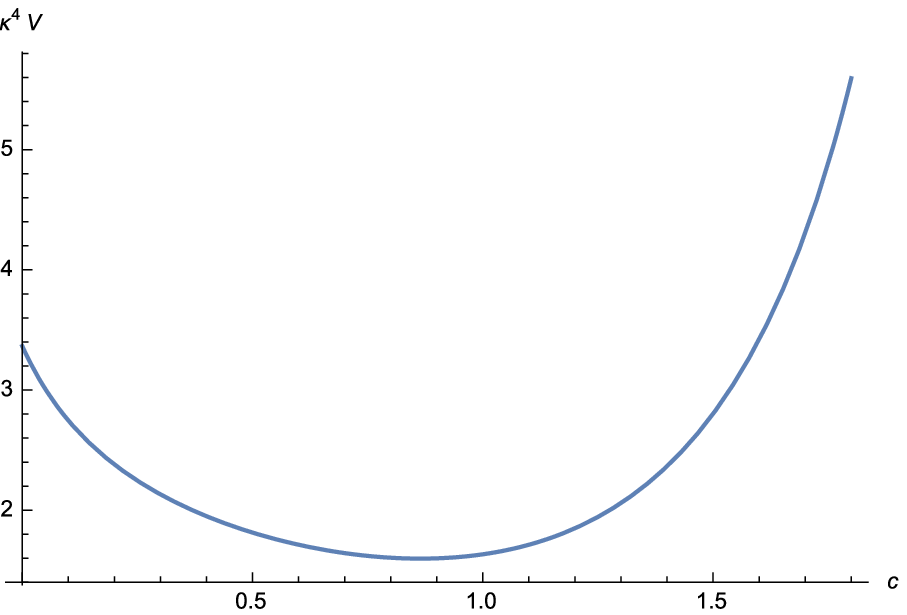}
        \caption{}
        \label{fig:IVa}
    \end{subfigure}
    ~ 
    \begin{subfigure}[b]{0.4\textwidth}
        \includegraphics[width=\textwidth]{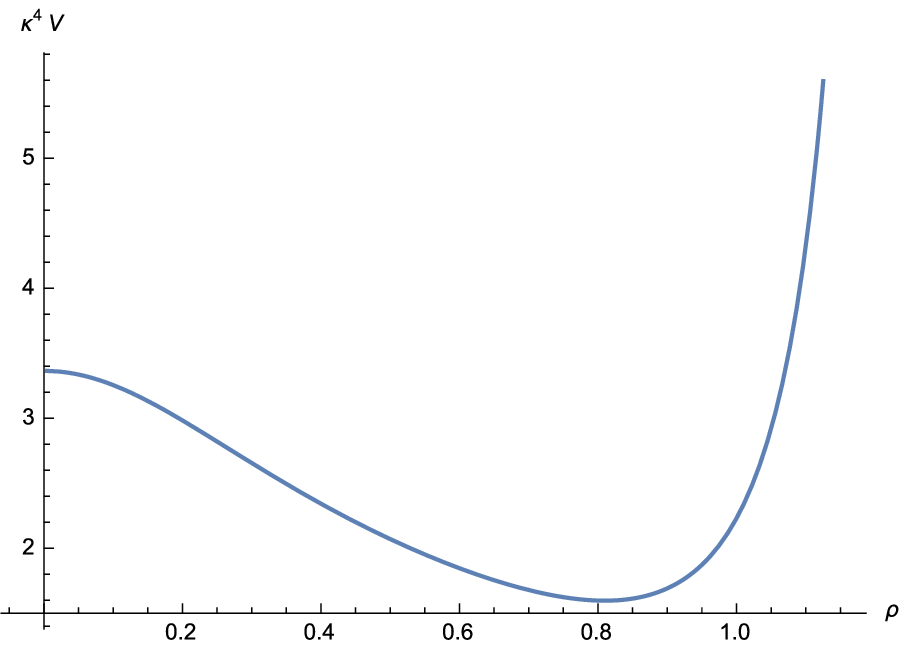}
        \caption{}
        \label{fig:IVb}
    \end{subfigure}
    \\ 
    \begin{subfigure}[b]{0.4\textwidth}
        \includegraphics[width=\textwidth]{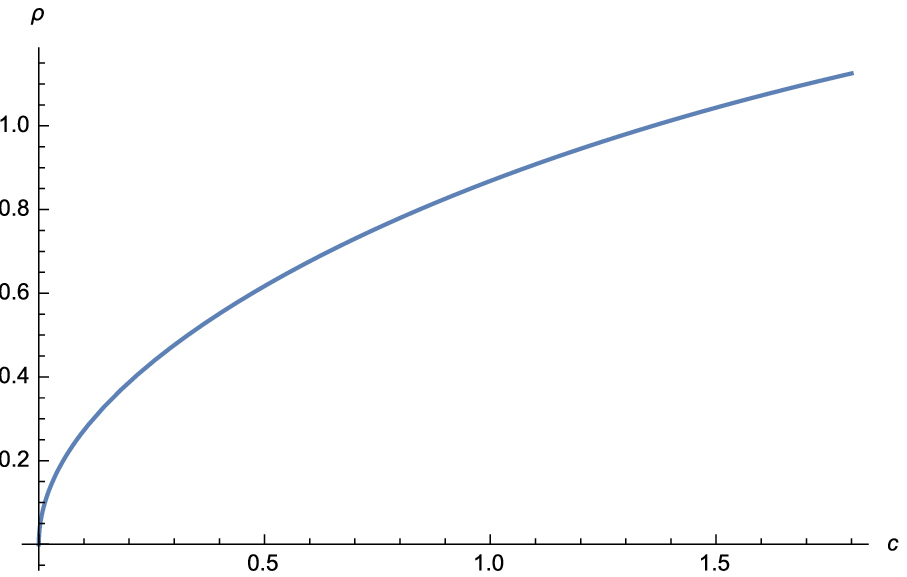}
        \caption{}
        \label{fig:IVc}
    \end{subfigure}
    \caption{The scalar potential for a model in \textbf{Region IV} of Fig.~\ref{plot01} with parameters chosen in (\ref{parameter:IV}) is plotted as a function of $c$ and $\rho$ coordinates in (a) and (b), respectively.  Plot in (c) shows the relation between $\rho$ and $c$.}\label{fig:RegionIV}
\end{figure}

We can choose for example
\begin{align}
v = 0.30, \quad x = 0.10, \quad y = 0.0, \quad m = 3.33.
\label{parameter:IV}
\end{align}
The scalar potential for these parameters as a function of $c$ and $\rho$ is plotted in Fig.~\ref{fig:IVa} and \ref{fig:IVb}, respectively.  The relation between $c$ and $\rho$ is shown in Fig.~\ref{fig:IVc} with physical domain $c > 0$.  For this choice of parameters, we have $\Delta = -0.99$.  However, it turns out that for the parameters given in (\ref{parameter:IV}), the constraint (\ref{determine_q}) only gives imaginary solutions for $q_-$.  This result also holds for any other set of parameters $v, x, y$ and $m$ in \textbf{Region IV}.   Therefore, we conclude that \textbf{Region IV} is unphysical.

\section{Conclusion}

In this paper we studied a generalised Fayet-Iliopoulos model based on a ${\rm U}(1)$ R-symmetry coupled to supergravity. Going to the Higgs phase in the limit of small supersymmetry breaking scale compared to the ${\rm U}(1)$ mass, we integrated out the massive vector multiplet and derived an effective field theory for the goldstino chiral multiplet characterised by a linear superpotential and an effective K\"ahler potential. By implementing the theory with a second gauged ${\rm U}(1)$ R-symmetry that remains spectator (and unbroken) in the above described Higgs phase of the first ${\rm U}(1)$, we were able to provide a microscopic model of inflation by supersymmetry breaking~\cite{Antoniadis:2017gjr}, upon identification of the inflaton with the goldstino superpartner having a dynamics driven by the effective field theory emerging from the integrating out procedure. The parameter space contains a region with a flat maximum at the origin where the second ${\rm U}(1)$ is unbroken and small field inflation takes place in agreement with CMB observations, until the inflaton rolls down to a `nearby' minimum having a tiny positive (tuneable) vacuum energy that can describe our observable universe.

In order to integrate out the heavy fields, we employed the formulation of supergravity with superconformal compensators in conformal superspace~\cite{Butter:2009cp}, to keep track of the normalisation of kinetic terms in the gravity multiplet and to facilitate the identification of the effective K\"ahler potential and superpotential.

It would be interesting to explore the possibility of realising our generalised Fayet-Iliopoulos model in a UV-complete theory, such as string theory with D-branes.

\section*{Acknowledgements}
This work was supported in part by the Swiss National Science Foundation, in part by the Labex ``Institut Lagrange de Paris'', in part by a CNRS PICS grant and in part by the ``CUniverse'' research promotion project by Chulalongkorn University (grant reference CUAASC).
The authors would like to thank Yifan Chen, Toshifumi Noumi and Ryo Yokokura for fruitful discussions.


\appendix

\section{Conformal supergravity and compensator superfields}
\label{app:conf-sugra}
We briefly review how to construct supergravity actions invariant under the diffeomorphisms and gauged superconformal transformations in a curved superspace
with coordinates $x^M=(x^M,\te^\mu,\te_\dmu)$, based on \cite{Butter:2009cp}. 

\subsubsection*{Gauging superconformal transformations}
Let us begin with the generators of superconformal transformations: translations, Lorentz transformations, dilatation, chiral $\rm U$(1) rotation, and special conformal transformations,
\begin{align}
\label{generators}
\hat P_A,\hat M_{ab},\hat D,\hat A,\hat K_A,
\end{align}
where the subscript $A$ in $\hat P_A,\hat K_A$ stand for the local Lorentz vector and spinor (undotted and dotted) indices $\{a,\al,\dal\}$. 
More concretely, the translation generators $\hat P_A$ are $(\hat P_a, \hat Q_\al, \hat{\bar Q}{}^\dal)$, and the special conformal transformation generators $\hat K_A$ are $(\hat K_a, \hat S_\al, \hat{\bar S}{}^\dal)$. The gauge transformation with parameter superfields $\bsxi^{\mathscr{A}}$ is generated by $\bsxi^{\mathscr{A}} \hat X_{\mathscr{A}}$, where $\hat X_{\mathscr{A}}$ represents the generators \eqref{generators}. Note also that calligraphic index such as $\mathscr{A}$ runs over the superconformal generators. We regard them as internal transformations, namely transformations acting on the local Lorentz coordinates.

To gauge them, we associate a gauge field $\bsh_M{}^{\mathscr{A}}$ with each generator,
\begin{align}
\bsh_M{}^{\mathscr{A}} = \bsh_M(\hat P)^A, ~ \bsh_M(\hat M)^{ab}, ~ \bsh_M(\hat D), ~ \bsh_M(\hat A), ~ \bsh_M(\hat K)^A.
\end{align} 
In particular, the gauge superfields $\bsh_M(\hat P)^A$, associated with the translations, are the vierbein superfields, which we will express using the ordinary symbol $\bsE_M{}^A$,
\begin{align}
\bsE_M{}^A = \bsh_M(\hat P)^A,
\end{align}
with its inverse $\bsE_A{}^M$. The (bosonic) vierbein and the gravitino are defined as the lowest components of $\bsE_m{}^A$,
\begin{align}
\bsE_m{}^a| = e_m{}^a, \qquad 
\bsE_m{}^\al|=\tfrac{1}{2}\psi_m{}^\al, \qquad 
\bsE_m{}^\dal|=\tfrac{1}{2}\bar\psi_m{}^\dal.
\end{align}

\subsubsection*{Covariant derivatives and curvatures}
Since we work in a curved superspace, it is more convenient to use the parallel transport generators $\nabla_A$, such that the generators $\nabla_A,\hat M_{ab},\hat D,\hat A,\hat K_A$ satisfying the commutation relations of the superconformal algebra except for $[\nab_A,\nab_B]$, which is given by\footnote{The commutation relation becomes the anti-commutator when the indices $A,B$ are both spinorial.} 
\begin{align}
[\nab_A, \nab_B] &= -\bsR_{AB}{}^{\mathscr{C}} \hat X_{\mathscr{C}} \\
&= - \bsR_{AB}(\nab)^C\nab_C - \tfrac{1}{2}\bsR_{AB}(\hat M)^{cd}\hat M_{dc} \nn\\
&{} \qquad - \bsR_{AB}(\hat D)\hat D - \bsR_{AB}(\hat A)\hat A - \bsR_{AB}(\hat K)^C\hat K_C,
\end{align}
where $\bsR_{AB}(\bullet)^{\mathscr{C}}$ are the curvature superfields, playing a role of the field-dependent structure ``constants''. 
Such parallel transport generators $\nab_A= \bsE_A{}^M\nab_M$ may be implemented with the gauge fields,\footnote{These relate the diffeomorphisms with parameters $\bsxi^M$ and the local Lorentz translations with parameters $\bsxi^A=\bsxi^M\bsE_M{}^A$. }
\begin{align}
\bsxi^M\nab_M\bsPhi = \cL(\bsxi^M\pd_M)\bsPhi - \bsxi^M \bsh_M{}^{\mathscr{A}'}\hat X_{\mathscr{A}'}\bsPhi,
\label{nabPhi}
\end{align}
where $\bsxi^M$ is a parameter superfield, $\cL(\xi^M\pd_M)$ is the Lie derivative, and the primed index $\mathscr{A}'$ indicates the generators except for the parallel transport ones. In particular, $\nab_M$ acts on a superfield $\bsPhi$ only with local Lorentz indices or with no indices as\footnote{Since the Lie derivative involves derivatives of the parameter $\bsxi$, it may be confusing to separate $\bsxi^M$ from $\nab_M$ in general. However, when $\bsPhi$ contains only lolcal Lorentz indices or no indices, the Lie derivative does not involve derivatives of $\bsxi$, and  makes sense to separate $\bsxi^M$ as in \eqref{nabPhi_localLorentz}.}
\begin{align}
\nab_M\bsPhi = \pd_M\bsPhi - \bsh_M{}^{\mathscr{A}'}\hat X_{\mathscr{A}'}\bsPhi,
\label{nabPhi_localLorentz}
\end{align}
which is nothing but the covariant derivative.

Now, our first goal is to construct actions which are invariant under the parallel transport $\nab_A$ and the other generators  $\hat M_{ab},\hat D,\hat A,\hat K_A$.\footnote{This goal boils down to the standard construction of non-supersymmetric gravitational theories, if we replace the superconformal generators by the Poincar\'e ones $\hat P_a,\hat M_{ab}$ and introduce the parallel transport $\nab_a$ for $\hat P_a$.}
It is clear from the definition of the parallel transport \eqref{nabPhi} that theories constructed in this way are invariant under the diffeomorphisms generated by the Lie derivatives.

\subsubsection*{Chiral and primary superfields}
Here, we introduce two important classes of superfields in conformal superspace. A chiral (anti-chiral) superfield is defined by $\bar\nab{}^\dal\bsPhi=0$ ($\nab_\al\bsPhi=0$), respectively. A primary superfield of weights $(\de,w)$ is defined by\footnote{For convenience, we present the commutation relations among $\nab,\hat D,\hat A$,
\begin{alignat}{3}
[\hat D,\nab_a]&=\nab_a, &\qquad
[\hat D,\nab_\al]&=\tfrac{1}{2}\nab_\al, &\qquad 
[\hat D,\bar\nab^\dal]&=\tfrac{1}{2}\bar\nab^\dal, \\
[\hat A,\nab_a]&=0, &\qquad
[\hat A,\nab_\al]&=-i\nab_\al, &\qquad 
[\hat A,\bar\nab^\dal]&=+i\bar\nab^\dal.
\end{alignat}}
\begin{align}
\hat D\bsPhi=\de\bsPhi, \qquad \hat A\bsPhi=iw\bsPhi, \qquad \hat K_A\bsPhi=0.
\end{align}

\subsubsection*{Curvature constraints}
Conformal supergravity imposes the following curvature constraints,
\begin{align}
&{} \bsR_{\al\bt}(\hat X)^{\mathscr{C}} = \bsR_{\dal\dbt}(\hat X)^{\mathscr{C}} = \bsR_{\al\dbt}(\hat X')^{\mathscr{C}} = 0, \\
&{} \bsR_{\al\dbt}(\nab)^c = 2i(\sig^c)_{\al\dbt}, \qquad \bsR_{\al\dbt}(\nab)^\ga = \bsR_{\al\dbt}(\nab)^\dga = 0, 
\label{const-nab1} \\
&{} \bsR_{\al b}(\nab)^{\mathscr{C}} = \bsR_{\dal b}(\nab)^{\mathscr{C}} = 0, 
\label{const-nab2}\\
&{} \bsR_{\al b}(\hat D) = \bsR_{\al b}(\hat A) = \bsR_{\dal b}(\hat D) = \bsR_{\dal b}(\hat A) = 0,
\end{align}
where $\hat X$ denotes all generators and $\hat X'$ the generators except the parallel transport. One can show that all unconstrained curvatures can be written in totally symmetric multi-spinor superfields $\boldsymbol{W}_{\al\bt\ga}$ and $\ol{\boldsymbol W}{}_{\dal\dbt\dga}$ which are chiral primary of weights $(3/2,1)$ and anti-chiral primary of weights $(3/2,-1)$, respectively.

\subsubsection*{Invariant actions}
As in the globally supersymmetric case, we will work with invariant actions which are classified into D-type and F-type. The D-type action takes the following form,
\begin{align}
S_D = \int \! d^4xd^4\te \, \bsE\mathbb{V},
\end{align}
where $\bsE$ is the super-determinant of the vierbein $\bsE_M{}^A$, and $\mathbb{V}$ is a scalar real primary superfield of weights $(2,0)$. On the other hand, the F-type action takes the following form,
\begin{align}
S_F = \int \! d^4xd^2\te \, \boldsymbol\cE \mathbb{W},
\end{align}
where $\boldsymbol\cE$ is the super-determinant of a part of the vierbein $\bsE_{M'}{}^{A'}$ with coordinate indices $M'=(m,\mu)$ and local Lorentz ones $A'=(a,\al)$, and $\mathbb{W}$ is a scalar chiral primary superfield of weight $(3,2)$.
A D-type action can be rewritten as an F-type one as \cite{Butter:2009cp}
\begin{align}
\int \! d^4xd^4\te \, \bsE\mathbb{V} = 
-\frac{1}{8}\int \! d^4xd^2\te \, \boldsymbol\cE\bar\nab^2\mathbb{V} - \frac{1}{8}\int \! d^4xd^2\bar\te \, \ol{\boldsymbol\cE}\nab^2\mathbb{V},
\label{DtoF}
\end{align}
where $\nab^2=\nab^\al\nab_\al$. It can also be shown that the two terms on the right hand side are actually equal.
On the other hand, an F-type action can also be rewritten as a D-type one as \cite{Butter:2009cp}
\begin{align}
\int \! d^4xd^2\te \, \boldsymbol\cE \mathbb{W}
= \int \! d^4xd^4\te \, \bsE\frac{\mathbb{W}\bsT}{-\frac{1}{4}\bar\nab^2\bsT},
\label{FtoD}
\end{align}
where $\bsT$ is an arbitrary superfield.

\subsubsection*{Compensators and gauge fixing}
To obtain supergravity theories which are super-Poincar\'e invariant, it is convenient to introduce compensator superfields and then fix them to break the $D,A,K$ gauge invariances. In this article, we introduce two compensator superfields $\bsC,\ol \bsC$, which are chiral primary of weights $(1,2/3)$ and anti-chiral primary of weights $(1,-2/3)$, respectively. 

Let us see a simple theory with one scalar chiral primary superfield $\bsPhi$ of weights $(0,0)$. A general invariant action may read
\begin{align}
S &= \kappa^{-3} \int \! d^4xd^2\te ~ \boldsymbol{\cE}\bsC^3 \cW(\bsPhi) + {\rm h.c.}
\nn\\
&\quad
- 3\kappa^{-2} \int \! d^4xd^4\te \, 
\bsE\bsC\ol \bsC e^{-\kappa^2 \cK(\bsPhi,\ol\bsPhi)/3},
\end{align}
where $\cW$ is the superpotential, which is real chiral primary of weights $(0,0)$, and $\cK$ is the K\"ahler potential, which is real primary of weights $(0,0)$. 

The next task to fix the gauge degrees of freedom of the dilatation, chiral ${\rm U}(1)$ and special conformal transformations. For this, we impose two conditions. One is 
\begin{align}
\bsh_M(\hat D)=0.
\label{Kfixing}
\end{align}
This completely exhausts the special conformal gauge degrees of freedom.
Combining this with the curvature constraints fixes the special conformal superfield $\bsh_A(\hat K)^B$. In particular, its spinor-spinor components are fixed to the forms
\begin{align}
& \bsh_\al(\hat K)_\bt=-\ep_{\al\bt}\ol \bsR, \qquad \bsh_\dal(\hat K)_\dbt=\ep_{\dal\dbt}\bsR, \label{hK-R} \\
& \bsh_\al(\hat K)_\dbt=-\bsh_\dbt(\hat K)_\al=-\tfrac{1}{2}\bsG_{\al\dbt}, \label{hK-G}
\end{align}
where in \eqref{hK-G}, the first equality is a nontrivial consequence of the condition \eqref{Kfixing} and the superfield $\bsG_{\al\dbt}$ just redefines $\bsh_\al(\hat K)_\dbt$.
The second gauge fixing condition is to set the compensators $\bsC,  \ol\bsC$ to some specific superfields so that this exhausts the dilatation and chiral ${\rm U}(1)$ gauge degrees of freedom. One easy choice is 
\begin{align}
\bsC=\ol \bsC=1.
\end{align}
This fixes the chiral ${\rm U}(1)$ gauge field $\bsh_B(\hat A)$ to
\begin{align}
\bsh_\al(\hat A)=\bsh_\dal(\hat A)=0, \qquad \bsh_{\al\dal}(\hat A)=-\tfrac{3}{2}\bsG_{\al\dal}.
\end{align}
The spinor covariant derivatives $\nab_\al,\bar\nab_\dal$ and thus the action then boil down to the standard Poincar\'e supergravity action with matter superfield $\bsPhi$ in \cite{Wess:1992cp}. Note however that the action has non-canonical kinetic terms in the gravity multiplet.

Alternatively, the canonically normalised kinetic terms in the gravity multiplet are realised by the following gauge condition
\begin{align}
\bsC=\ol \bsC=e^{-\kappa^2\cK/6}.
\label{BDfixing-can}
\end{align}
In contrast with the other fixing above, this fixes the chiral ${\rm U}(1)$ rotation gauge field $\bsh_B(\hat A)$ to non-zero components determined by the K\"ahler potential $\cK$, which are called K\"ahler connections \cite{Binetruy:2000zx,Butter:2009cp}. 

Note that $\bsR$ ($\ol\bsR$) is chiral (anti-chiral) with respect to the gauge-fixed covariant derivatives $\cD_\al$ ($\bar\cD_\dal$), which are obtained from $\nab_\al$ ($\bar\nab_\dal$) by setting $\bsh(\hat D)=0$ and replacing $\bsh(\hat A),\bsh(\hat K)$ by their gauge-fixed forms.
For instance, this replacement converts the chiral projector as
\begin{align}
-\tfrac{1}{4}\bar\nab^2\bsPhi\big|_{\rm gauge\,\,fixing} = -\tfrac{1}{4}(\bar\cD^2-8\bsR)\bsPhi,
\label{chiralproj}
\end{align}
where $\bsPhi$ is a primary superfield of weights $(0,0)$. One can show \cite{Butter:2009cp} that theories after the gauge fixing \eqref{Kfixing} and \eqref{BDfixing-can} are expressed in terms of $\cD_A,\bsR,\bsG,\boldsymbol{W},\ol{\boldsymbol W}$.

\subsubsection*{Relation to other formulations}
Here, we comment on the relation to other formulations.
We have already mentioned above that the gauge fixing by \eqref{Kfixing} and \eqref{BDfixing-can} gives supergravity theories in the K\"ahler superspace. The superfields $\bsR,\ol\bsR,\bsG_{\al\dal}$ above correspond to those of the formulation in \cite{Binetruy:2000zx}. The covariant derivatives for the K\"ahler  superspace are different from those for the superspace in Wess and Bagger \cite{Wess:1992cp} by the gauge fixed $\bsh_M(\hat A)$.  
It is possible to convert the K\"ahler superapce to the superspace of Wess and Bagger by redefining the torsion components \cite{Binetruy:2000zx} to eliminate the remnant $\bsh_M(\hat A)$. On the other hand, \cite{Kugo:2016zzf,Kugo:2016lum} showed that the superconformal tensor calculus \cite{Freedman:2012zz,Kugo:1982mr,Kugo:1983mv} is obtained by fixing the gauge degrees of freedom with all $\te$-components of the gauge parameter superfield $\bsxi^{\mathscr{A}}$ except the lowest ones $\bsxi^{\mathscr{A}}|$.

Whichever formulation we adopt, the scalar potential takes the same form, given by the following standard formula,
\begin{align}
\kappa^4 \cV = e^{\ka^2\cK}(g^{\Phi\bar\Phi}D_\Phi \cW D_{\bar\Phi}\ol \cW - 3\ka^2\cW\ol \cW),
\label{scpot}
\end{align}
where $D_\Phi \cW=\pd_\Phi \cW+\ka^2(\pd_\Phi \cK)\cW$ and $g^{\Phi\bar\Phi}=(\pd_\Phi\pd_{\bar\Phi}\cK)^{-1}$.

\subsubsection*{Proof of the identity \eqref{WWtoWDV}}
Below, we give an outline of proving the identity \eqref{WWtoWDV}. We first show using the chirality of $\bsW^\al$ that
\begin{align}
\int d^4xd^2\te \, \boldsymbol\cE \bsW^\al \bsW_\al 
= -\frac{1}{4}\int d^4xd^2\te \, \boldsymbol\cE \bar\nab^2(\bsW^\al\nab_\al \bsV).
\end{align}
Applying the conversion formula \eqref{DtoF} to this integral plus its Hermitian conjugate gives the integral 
\begin{align}
\frac{1}{4} \int d^4xd^2\te \, \boldsymbol\cE \bsW^\al \bsW_\al + {\rm h.c.} = \frac{1}{2}\int d^4xd^4\te \, \bsE\bsW^\al\nab_\al \bsV.
\label{WWtoWDV1}
\end{align}
The next step is to integrate the right hand side by part to put $\nab_\al$ on $\bsW^\al$. This is straightforward in the globally supersymmetric case, but not in the present case. One reason is that the statement that a total derivative vanishes in an integral over the superspace is correct when the total derivative is with respect to a coordinate, not to a local Lorentz one. Another reason is that the integrand involves the vierbein. Taking them into account, we consider the following total derivative,
\begin{align}
\pd_M(\bsE\bsE_\al{}^M\bsW^\al \bsV).
\end{align}
This vanishes in the integral $\int d^4xd^4\te$. We can then prove that the derivative can be replaced by the covariant derivative\footnote{To prove this, we need the definition of the gauge transformation of the vierbein with parameter $\bsxi$ for generators other than the parallel transport,
\begin{align}
\bsxi^{\mathscr{B}'}X_{\mathscr{B}'}\bsE_M{}^A = - \bsE_M{}^C \bsxi^{\mathscr{B}'} [X_{\mathscr{B}'},\nab_C]^A,
\end{align}
where $[X_{\mathscr{B}'},\nab_C]^A$ picks up the coefficient of $\nab^A$ in $[X_{\mathscr{B}'},\nab_C]$.
}
\begin{align}
\pd_M(\bsE\bsE_\al{}^M\bsW^\al \bsV) 
&= \nab_M(\bsE\bsE_\al{}^M\bsW^\al \bsV) \nn\\
&= \nab_M(\bsE\bsE_\al{}^M)\bsW^\al \bsV + \bsE\nab_\al(\bsW^\al \bsV).
\end{align}
Let us focus on $\nab_M(\bsE\bsE_\al{}^M)$. We can actually prove the following identity
\begin{align}
\nab_M(\bsE\bsE_\al{}^M) = -\bsR_{B\al}(\nab)^B,
\end{align}
which vanishes thanks to the curvature constraints \eqref{const-nab1}, \eqref{const-nab2}. Combining these results, we find the desired identity
\begin{align}
0 = \int d^4xd^4\te \, \pd_M(\bsE\bsE_\al{}^M\bsW^\al \bsV) 
= \int d^4xd^4\te \, \bsE\nab_\al(\bsW^\al \bsV).
\end{align}
Applying this to the right hand side of \eqref{WWtoWDV1} gives the identity \eqref{WWtoWDV}.


\end{document}